\documentclass[submission,Phys]{SciPost}

\usepackage[utf8]{inputenc} 
\usepackage[T1]{fontenc} 	
\usepackage[english]{babel} 

\usepackage[bitstream-charter]{mathdesign}
\usepackage{geometry} 		
\usepackage{amsmath} 		
\usepackage{mathtools} 		
\usepackage{float} 			
\usepackage{graphicx} 		
\usepackage{tabularx} 		
\usepackage{booktabs} 		
\usepackage{color, xcolor} 	
\usepackage{pdfpages} 		
\usepackage{extarrows} 		
\usepackage{multirow} 		
\usepackage{multicol} 		
\usepackage[subrefformat=parens]{subcaption}
\usepackage{enumitem} 		
\usepackage{xspace} 		
\usepackage{stackrel} 		
\usepackage{tikz} 			
\usetikzlibrary{positioning,calc}
\usepackage{braket} 		
\usepackage{bm} 			
\usepackage{tensor} 		
\usepackage{slashed} 		
\usepackage{siunitx} 
\usepackage{lastpage} 		
\usepackage{cite} 			
\usepackage[normalem]{ulem} 
\usepackage{fontawesome} 	
\usepackage{tocloft} 		
\usepackage{titlesec} 		
\usepackage{doi} 			
\usepackage{hyperref} 		
\usepackage[most]{tcolorbox} 					
\usepackage[nameinlink, capitalize]{cleveref} 	
\usepackage[nottoc, notlot, notlof]{tocbibind} 	
\usepackage[ruled, vlined]{algorithm2e} 		
\usepackage{makecell}
\usepackage{setspace}

\makeatletter
\def\BState{\State\hskip-\ALG@thistlm}
\makeatother

\makeatletter
\@ifundefined{pdfoutput}{}{\DeclareGraphicsRule{*}{mps}{*}{}}
\makeatother

\makeatletter
\DeclareRobustCommand*{\bfseries}{%
   \not@math@alphabet\bfseries\mathbf
   \fontseries\bfdefault\selectfont
   \boldmath
}
\makeatother

\hypersetup{
	pdftitle={ML-MEM},
	pdfauthor={Heimel et al.},
	colorlinks=true, 			
	linkcolor={red!50!black}, 	
	citecolor={blue!50!black}, 	
	urlcolor={blue!80!black} 	
} 

\DeclareSymbolFont{usualmathcal}{OMS}{cmsy}{m}{n}
\DeclareSymbolFontAlphabet{\mathcal}{usualmathcal}

\SetArgSty{textnormal}
\SetKwComment{Comment}{{\small\#}~}{}
\SetCommentSty{mycommfont}

\setitemize{itemsep=0pt, parsep=0pt} 				
\setenumerate{itemsep=0pt, parsep=0pt} 				
\setitemize{itemsep=2pt,topsep=2pt,parsep=0pt,partopsep=0pt,leftmargin=*}
\setenumerate{itemsep=0pt,topsep=2pt,parsep=0pt,partopsep=0pt,labelindent=3pt,leftmargin=*}
\newcommand{\ie}{i.e.\@\xspace} 		

\newcommand{\Langle}{\bigl\langle}
\newcommand{\Rangle}{\bigr\rangle}
\newcommand{\XLangle}{\Bigl\langle}
\newcommand{\XRangle}{\Bigr\rangle}

\newcommand{\XXXLangle}{\Biggl\langle}
\newcommand{\XXXRangle}{\Biggr\rangle}

\newcommand{\qqquad}{\qquad\quad}

\newcommand{\normal}{\mathcal{N}}
\newcommand{\lag}{\mathscr{L}}

\newcommand{\loss}{\mathcal{L}} 	

\newcommand{\pl}{p_\text{latent}}
\newcommand{\pd}{p_\text{data}}
\newcommand{\pmd}{p_\theta}
\newcommand{\qmd}{q_\varphi}
\newcommand{\gmd}{J_\varphi}
\newcommand{\epsmd}{\epsilon_\psi}

\newcommand{\pytorch}{\textsc{PyTorch}\xspace}

\newcommand{\pythia}{\textsc{Pythia}\xspace}
\newcommand{\herwig}{\textsc{Herwig}\xspace}
\newcommand{\delphes}{\textsc{Delphes}\xspace}

\newcommand{\vegas}{\textsc{Vegas}\xspace}

\newcommand{\arXiv}[2][]{%
	\ifthenelse{\equal{#1}{}}%
	{\href{http://arxiv.org/abs/#2}{arXiv:#2}}%
	{\href{http://arxiv.org/abs/#2}{arXiv:#2~[#1]}}}

\def\slashchar#1{\setbox0=\hbox{$#1$}           
   \dimen0=\wd0                                 
   \setbox1=\hbox{/} \dimen1=\wd1               
   \ifdim\dimen0>\dimen1                        
      \rlap{\hbox to \dimen0{\hfil/\hfil}}      
      #1                                        
   \else                                        
      \rlap{\hbox to \dimen1{\hfil$#1$\hfil}}   
      /                                         
   \fi}

\newcommand{\tikznode}[2]{%
\ifmmode%
\tikz[remember picture,baseline=(#1.base),inner sep=0pt] \node (#1) {$#2$};%
\else
\tikz[remember picture,baseline=(#1.base),inner sep=0pt] \node (#1) {#2};%
\fi}

\def\mathswitchr#1{\relax\ifmmode{\text{#1}}\else$\text{#1}$\xspace\fi}
\def\mathswitch#1{\relax\ifmmode#1\else$#1$\xspace\fi}

\usetikzlibrary{arrows, shapes.geometric, arrows.meta, shapes, decorations.pathreplacing, fit}
\graphicspath{{./figs/}}
\begin{document}


\vspace*{-2.5em}
\hfill{}
\vspace*{0.5em}

\begin{center}{\Large \textbf{
Precision-Machine Learning for the Matrix Element Method 
}}\end{center}

\begin{center}
Theo Heimel\textsuperscript{1},
Nathan Huetsch\textsuperscript{1},
Ramon Winterhalder\textsuperscript{2},\\
Tilman Plehn\textsuperscript{1},
and Anja Butter\textsuperscript{1,3}
\end{center}

\begin{center}
{\bf 1} Institut für Theoretische Physik, Universität Heidelberg, Germany
\\
{\bf 2}
CP3, Universit\'e catholique de Louvain, Louvain-la-Neuve, Belgium
\\
{\bf 3} LPNHE, Sorbonne Universit\'e, Universit\'e Paris Cit\'e, CNRS/IN2P3, Paris, France
\end{center}


\vspace{-1cm}
\section*{Abstract}
{\bf The matrix element method is the LHC inference method of choice
  for limited statistics. We present a dedicated machine learning framework, based on efficient phase-space integration, a learned
  acceptance and transfer function. It is based on a choice of INN and diffusion networks, and a transformer to solve jet
  combinatorics. We showcase this setup for the
  CP-phase of the top Yukawa coupling in associated Higgs
  and single-top production.}

\vspace{10pt}
\noindent\rule{\textwidth}{1pt}
\tableofcontents\thispagestyle{fancy}
\noindent\rule{\textwidth}{1pt}

\clearpage

\section{Introduction and reference process}
\label{sec:intro}

Optimal analyses are the key challenge for the current and future LHC
program, including specific model-based as well 
simulation-based search strategies. A classic method is the matrix
element method (MEM), developed for the top physics program at the
Tevatron~\cite{Kondo:1988yd,Kondo:1991dw}. It derives its optimality
from the Neyman-Pearson lemma and the fact that all information for a
given hypothesis is encoded in the differential cross section. In the
MEM, we compute likelihood ratios for individual events, such that the
log-likelihood ratio of an event sample is the sum of the
event-wise log-likelihood ratios. A combination of events to a
kinematic distribution is not necessary~\cite{Cranmer:2006zs}.

The MEM was first used in the top mass
measurement~\cite{D0:1998eiz,Abazov:2004cs,CDF:2006nne,Fiedler:2010sg}
and the discovery of the single-top production
process~\cite{Giammanco:2017xyn} at the Tevatron. At the LHC, there
exist several
studies~\cite{Artoisenet:2010cn,Alwall:2010cq,Andersen:2012kn,Artoisenet:2013vfa,Englert:2015dlp,FerreiradeLima:2017iwx,Brochet:2018pqf}
and analysis
applications~\cite{CMS:2015enw,ATLAS:2015jmq,CMS:2015cal,Gritsan:2016hjl,FerreiradeLima:2017iwx}. The
critical challenge to MEM analyses is the integration over all
possible parton-level configurations which could lead to the analyzed
observed events. It can be solved by using modern machine learning (ML)
for a fast and efficient combination of simulation and
integration~\cite{Bury:2020ewi,Butter:2022vkj}.  A
related ML approach to likelihood extraction is the classifier-based estimation of likelihood ratios~\cite{Brehmer:2019xox}.

We present a comprehensive simulation and
integration framework for the MEM, based on modern machine learning~\cite{Butter:2022rso,Plehn:2022ftl}.  It makes extensive use of
generative networks, which are transforming LHC simulations and
analyses just like any other part of our lives. This starts with phase-space integration and
sampling~\cite{Chen:2020nfb,Bothmann:2020ywa,Gao:2020vdv,Gao:2020zvv,Danziger:2021eeg,Heimel:2022wyj}
and continues with more LHC-specific tasks like event
subtraction~\cite{Butter:2019eyo}, event
unweighting~\cite{Verheyen:2020bjw,Backes:2020vka}, loop integrations~\cite{Winterhalder:2021ngy}, or super-resolution enhancement~\cite{DiBello:2020bas,Baldi:2020hjm}. At the LHC, generative networks generally work in interpretable physics phase spaces, for
example scattering
events~\cite{dutch,gan_datasets,DijetGAN2,Butter:2019cae,Alanazi:2020klf,Butter:2021csz,Butter:2023fov},
parton
showers~\cite{locationGAN,Andreassen:2018apy,Bothmann:2018trh,Dohi:2020eda,Buhmann:2023pmh,Leigh:2023toe,Mikuni:2023dvk,Buhmann:2023zgc},
and detector
simulations~\cite{Paganini:2017hrr,deOliveira:2017rwa,Paganini:2017dwg,Erdmann:2018kuh,Erdmann:2018jxd,Belayneh:2019vyx,Buhmann:2020pmy,Buhmann:2021lxj,Krause:2021ilc,
  ATLAS:2021pzo,Krause:2021wez,Buhmann:2021caf,Chen:2021gdz,
  Mikuni:2022xry,ATLAS:2022jhk,Krause:2022jna,Cresswell:2022tof,Diefenbacher:2023vsw,
  Hashemi:2023ruu,Xu:2023xdc,Diefenbacher:2023prl,Buhmann:2023bwk,Buckley:2023rez,Diefenbacher:2023flw
}. These networks can be trained on first-principle simulations and
are easy to handle, efficient to ship, powerful in amplifying the
training samples~\cite{Butter:2020qhk,Bieringer:2022cbs}, and --- most
importantly --- precise~\cite{Butter:2021csz, Winterhalder:2021ave,Nachman:2023clf,Das:2023ktd}. Conditional versions of
these established generative networks then enable new analysis methods, like
probabilistic
unfolding~\cite{Datta:2018mwd,Bellagente:2019uyp,Andreassen:2019cjw,Bellagente:2020piv,Backes:2022vmn,Leigh:2022lpn,Shmakov:2023kjj,Ackerschott:2023nax,Diefenbacher:2023wec},
inference~\cite{Bieringer:2020tnw,Butter:2022vkj}, or anomaly
detection~\cite{Nachman:2020lpy,Hallin:2021wme,Raine:2022hht,Hallin:2022eoq,Golling:2022nkl,Sengupta:2023xqy}.

We introduce a new MEM-ML-analysis framework in Sec.~\ref{sec:memml_setup}. It combines two
generative network and one classifier network and pushes the precision
beyond our conceptual study~\cite{Butter:2022vkj}, towards an
experimentally required level. For a fast and bi-directional evaluation
we use the established cINNs with advanced coupling
layers~\cite{Butter:2021csz}, updated to current precision
requirements in Sec.~\ref{sec:memml_improve}. In Sec.~\ref{sec:memml_acc}, we add a learned 
acceptance network. In Sec.~\ref{sec:diffusion}, we show how a generative diffusion
network~\cite{Butter:2023fov} improves the precision, albeit at the
expense of speed. Finally, we employ a transformer 
architecture~\cite{Mikuni:2020wpr,Butter:2023fov,Finke:2023veq} to solve the
jet combinatorics in Sec.~\ref{sec:tramsformer}.
This series of improvements allows us to extract likelihood 
distributions from small and moderate-size event samples without 
a network bias and with close-to-optimal performance.

\subsubsection*{Reference process}

The focus of this paper is entirely on our new ML-method to enable MEM
analyses at the LHC. However, we use an established, challenging, and
realistic physics process to illustrate our method. This reference
process is introduced and discussed in Ref.~\cite{Butter:2022vkj}. We
target the purely hadronic signature
\begin{align}
  p p \to t H j \to (b jj) \; (\gamma \gamma) \; j + \text{jets}\; ,
  \label{eq:proc}
\end{align}
with up to four additional jets from QCD radiation.
The production process allows for a measurement of a CP-phase in the
top Yukawa coupling at future LHC
runs~\cite{Buckley:2015vsa,Ren:2019xhp,Bortolato:2020zcg,Bahl:2020wee,Martini:2021uey,Goncalves:2021dcu,Barman:2021yfh,Bahl:2021dnc,Kraus:2019myc}. The
challenge of having to work with small event numbers is motivated by
choosing the rare decay channel $H \to \gamma \gamma$, which allows us
to control continuum backgrounds efficiently. The total cross
section is 43.6~fb, when we combine top and anti-top production.

To probe the symmetry structure of the Yukawa coupling, we introduce a mixed
CP-even and CP-odd
interaction~\cite{Artoisenet:2013puc},
\begin{align}
  \lag_{t\bar{t}H} = - \frac{y_t}{\sqrt{2}}
  \Big[ a\cos \alpha \; \bar{t}t + 
  ib\sin \alpha \; \bar{t}\gamma_5 t \Big] H \; .
\label{eq:LttH}
\end{align}
Choosing $a = 1$ and $b=2/3$~\cite{Demartin:2015uha} keeps the cross
section for $gg\to H$ constant.  The model parameter we target with
the matrix element method is the CP-angle $\alpha$. For more details
on this reference process we refer to our conceptual
study~\cite{Butter:2022vkj}. Obviously, all our findings can be generalized
to other LHC processes.

\section{ML-matrix element method}
\label{sec:memml_setup}

The matrix element method is a simulation-based inference method which
uses the fact that for a given parameter of interest, $\alpha$, the
likelihood can be extracted from a simulation of the differential
cross section. It describes the hard scattering process and 
factorizes into the total cross section and a normalized probability
density,
\begin{align}
 \frac{d \sigma(\alpha)}{d x_\text{hard}}= \sigma(\alpha)\; p(x_\text{hard}|\alpha)
 \qquad \Leftrightarrow \qquad
 p(x_\text{hard}|\alpha) = \frac{1}{\sigma(\alpha)} \;
 \frac{d \sigma(\alpha)}{d x_\text{hard}} \; .
\label{eq:digma_diff_a}
\end{align}
Given the hard process, we then simulate the parton shower,
hadronization, detector effects, and the reconstruction of analysis
objects, with a forward-transfer or response function $r$~\cite{Cowan:1998ji}. 
This function is assumed to be independent of the theory parameter $\alpha$
\begin{align}
\begin{split}
\begin{tikzpicture}[node distance=-0.2cm and 3cm]
\node (A) 
    {$x_\text{hard}$};
\node[above right=of A] (B) 
    {$x_\text{reco}$};
\node[below right=of A] (C)    
    {rejected};
    \draw[-stealth] (A) -- ( $ (A.0)!0.2!(B.west|-A.0) $ ) |- (B.west) node[above,pos=0.75] {${\scriptstyle r(x_\text{reco}|x_\text{hard})}$};
    \draw[-stealth] (A) -- ( $ (A.0)!0.2!(C.west|-A.0) $ ) |- (C.west) node[below,pos=0.75] {${\scriptstyle p_\text{reject}(x_\text{hard})}$};
\end{tikzpicture} \; .
\end{split}
\end{align}
The detector geometry and acceptance cuts will lead to, either, a
valid reco-level event $x_\text{reco}$ or a rejected event, introducing 
$p_\text{reject}(x_\text{hard})$ as the probability that a given hard event $x_\text{hard}$ is rejected.
The transfer function $r$ is not 
normalized, and a proper normalization condition defines the efficiency or acceptance function,
\begin{align}
  \epsilon(x_\text{hard}):=\int d x_\text{reco} \;r(x_\text{reco}|x_\text{hard})
  = 1  - p_\text{reject}(x_\text{hard})\; .
  \label{eq:def_eff}
\end{align}
Using the transfer function we can parametrize the forward evolution of the differential cross section following
\begin{align}
    \frac{d \sigma_\text{fid}(\alpha)}{d x_\text{reco}} &= \int d x_\text{hard}\; 
    r(x_\text{reco}|x_\text{hard}) \;
    \frac{d \sigma(\alpha)}{d x_\text{hard}} \; ,
    \label{eq:full_forward}
\end{align}
where the subscript `fid' indicates that the reco-level phase space is different from the parton level. 
In this relation we can use Eq.\eqref{eq:def_eff} to
replace $r$ with a normalized transfer probability $p(x_\text{reco}|x_\text{hard})$,
\begin{align} 
  r(x_\text{reco}|x_\text{hard}) = \epsilon(x_\text{hard}) \; p(x_\text{reco}|x_\text{hard}) \qquad
  \text{with} \qquad
  \int d x_\text{reco}\; p(x_\text{reco}|x_\text{hard}) = 1 \; .
  \label{eq:forward_to_tf}
\end{align}
Inserting Eq.\eqref{eq:forward_to_tf} in Eq.\eqref{eq:full_forward} we obtain the final expression for the differential cross section
\begin{align}
    \frac{d \sigma_\text{fid}(\alpha)}{d x_\text{reco}} 
    &= \int d x_\text{hard}\; 
    \epsilon(x_\text{hard}) \; p(x_\text{reco}|x_\text{hard}) \;
    \frac{d \sigma(\alpha)}{d x_\text{hard}} \; ,
    \label{eq:full_forward_2}
\end{align}
Equivalent to Eq.\eqref{eq:digma_diff_a} we can now define the likelihood for reco-level events in terms of the fiducial cross section and the differential cross section
\begin{align}
  \frac{d \sigma _\text{fid}(\alpha)}{d x_\text{reco}}= \sigma_\text{fid}(\alpha)\; p(x_\text{reco}|\alpha)
 \qquad \Leftrightarrow \qquad
  p(x_\text{reco}|\alpha)
  = \frac{1}{\sigma_\text{fid}(\alpha)} \frac{d \sigma_\text{fid}(\alpha)}{d x_\text{reco}}\;.
  \label{eq:reco_like}
\end{align}
To obtain the fiducial cross section $\sigma_\text{fid}(\alpha)$, we now need to integrate Eq.\eqref{eq:full_forward_2} over the reco-level phase space 
\begin{align}
  \sigma_\text{fid}(\alpha)
  &=\int d x_\text{reco}\int d x_\text{hard}\; 
   \epsilon(x_\text{hard}) \; p(x_\text{reco}|x_\text{hard})\;
  \frac{d \sigma(\alpha)}{d x_\text{hard}} \notag\\
   &=\int d x_\text{hard}\; 
   \epsilon(x_\text{hard}) \; 
  \frac{d \sigma(\alpha)}{d x_\text{hard}} \notag\\
  &=\sigma(\alpha)
  \int d x_\text{hard}\; 
    \epsilon(x_\text{hard}) \;
    p(x_\text{hard}|\alpha) \notag \\
  &= \sigma(\alpha)\;\Langle \epsilon(x_\text{hard}) \Rangle_{x\sim p(x_\text{hard}|\alpha)}\; ,
  \label{eq:sigma_fid}
\end{align}
where we first use Eq.\eqref{eq:forward_to_tf} to integrate out the reco-level phase space and then replace the differential cross section using Eq.\eqref{eq:digma_diff_a}. This allows us to express the integral in terms of the 
average acceptance
$\langle\epsilon\rangle_\alpha$ which is used to evaluate the integral numerically.
Using Eq.~\eqref{eq:full_forward_2} in Eq.~\eqref{eq:reco_like} we obtain the
final expression for the reco-level likelihood
\begin{align}
  p(x_\text{reco}|\alpha)
  &= \frac{1}{\sigma_\text{fid}(\alpha)} \int d x_\text{hard}\; \frac{d \sigma(\alpha)}{d x_\text{hard}} \; \epsilon(x_\text{hard}) \;
  p(x_\text{reco}|x_\text{hard}) \; .
  \label{eq:likeli_3} 
\end{align}
Note that in our training dataset, consisting of simulated event pairs $(x_\text{reco}, x_\text{hard})$, the hard-scattering momenta are not distributed according to Eq.\eqref{eq:digma_diff_a}, because it does not contain events $x_\text{hard}$ that have been rejected. Consequently, the accepted $x_\text{hard}$ are distributed as
\begin{align}
 p_\text{fid}(x_\text{hard}|\alpha) = \frac{1}{\sigma_\text{fid}(\alpha)} \;
 \frac{d \sigma(\alpha)}{d x_\text{hard}} \;
 \epsilon(x_\text{hard}) \;.
 \label{eq:p_fid_hard}
\end{align}
This means, we can directly relate the reco-level likelihood to a modified parton-level likelihood
\begin{align}
  p(x_\text{reco}|\alpha)
  &= \int d x_\text{hard}\;  p(x_\text{reco}|x_\text{hard}) \; p_\text{fid}(x_\text{hard}|\alpha)\;,
  \label{eq:likeli_prob} 
\end{align}
which connects the MEM with the completeness relation from statistics.

\subsubsection*{Acceptance classifier and transfer network}

To compute the reco-level likelihood defined in Eq.\eqref{eq:likeli_3}
we rely on $\epsilon(x_\text{hard})$ and
$p(x_\text{reco}|x_\text{hard})$, defined through a forward
simulation. We encode both functions in neural networks trained on these forward simulations. 

First, the acceptance $\epsilon(x_\text{hard})$
can be encoded as a standard classifier network
\begin{align}
x_\text{hard}\xrightarrow{\text{Acceptance network}}\epsmd(x_\text{hard})  \;,
\end{align}
where $\psi$ denotes the trainable network parameters.
Given the input
$x_\text{hard}$ it learns the labels $1$ for accepted events and $0$
otherwise. Because the network is a classifier with a cross entropy
loss, its output will be the acceptance probability for the given
event.

The transfer probability introduced in Eq.\eqref{eq:forward_to_tf} is encoded in a
generative network with density estimation capability, like a normalizing flow
or diffusion model, and is trained on
event pairs $(x_\text{reco}, x_\text{hard})$. For this training
dataset, we only include accepted events. The generative network
defines a bijective mapping between Gaussian random numbers and
reco-level phase space conditioned on parton-level events,
\begin{align}
  x_\text{reco} \sim \pmd (x_\text{reco}|x_\text{hard})
  \quad 
  \xleftrightarrow{\text{Transfer network}}
  \quad
  r \sim \pl(r) \; ,
  \label{eq:transfer_cinn}
\end{align}
with trainable parameters $\theta$.
This mapping can than be used for density estimation in the forward
direction and for conditional generation of reco-level events in the
inverse direction.

\subsubsection*{Sampling-cINN}

\begin{figure}[t]
\centering
\definecolor{Rcolor}{HTML}{E99595}
\definecolor{Gcolor}{HTML}{C5E0B4}
\definecolor{Bcolor}{HTML}{9DC3E6}
\definecolor{Ycolor}{HTML}{FFE699}

\tikzstyle{expr} = [circle, minimum width=1.8cm, minimum height=1.8cm, text centered, align=center, inner sep=0, fill=Ycolor, draw]
\tikzstyle{txt_huge} = [align=center, font=\Huge, scale=2]
\tikzstyle{txt} = [align=center, font=\LARGE]
\tikzstyle{cinn} = [double arrow, double arrow head extend=0cm, double arrow tip angle=130, shape border rotate=90, inner sep=0, align=center, minimum width=2.1cm, minimum height=2.3cm, fill=Gcolor, draw]
\tikzstyle{cinn_black} = [cinn, minimum height=2.5cm, fill=black]
\tikzstyle{network_black} = [cinn, minimum height=2.1cm, fill=black]
\tikzstyle{network} = [rectangle, align=center, rounded corners=0.3ex, minimum width=2.3cm, minimum height=1.6cm, draw]
\tikzstyle{arrow} = [thick,-{Latex[scale=1.0]}, line width=0.2mm, color=black]
\tikzstyle{line} = [thick, line width=0.2mm, color=black]

\begin{tikzpicture}[node distance=2.4cm, scale=0.7, every node/.style={transform shape}]

\node (likeli) [txt] {$p(x_\text{reco}|\alpha)\; = $};
\node (sigfid) [expr, right of=likeli, xshift = 0.4cm] {\LARGE $\frac{1}{\sigma_\text{fid}}$};
\node (langle) [txt_huge, right of=sigfid, xshift = -1.6cm] {$\langle$};
\node (jac) [expr, right of=langle, xshift = -0.6cm] {$\dfrac{1}{q(x_\text{hard})}$};
\node (dcs) [expr, right of=jac] {\LARGE $\frac{d \sigma(\alpha)}{d x_\text{hard}}$};
\node (acc) [expr, right of=dcs] {$\epsilon(x_\text{hard})$};
\node (dens) [expr, right of=acc] {$p(x_\text{reco}|$\\$x_\text{hard})$};
\node (langle) [txt_huge, right of=dens, xshift = -1.5cm] {$\rangle$};

\node (unfcinn_b) [cinn_black, above of=jac, yshift=3.1cm] {};
\node (unfcinn) [cinn, above of=jac, yshift=3.1cm, fill=Bcolor] {Sampling\\cINN};
\node (tracinn_b) [cinn_black, above of=dens, yshift=3.1cm] {};
\node (tracinn) [cinn, above of=dens, yshift=3.1cm] {Transfer\\network};

\node (accnet) [network, above of=acc, yshift=.8cm, fill=Rcolor] {Acceptance\\network};

\node (alpha) [txt, above of=likeli, yshift=2.9cm] {$\alpha$};
\node (xr) [txt, above of=alpha, yshift=1.4cm] {$x_\text{reco}$};
\node (random) [txt, above of=unfcinn, yshift=0.3cm] {\normalsize$\{r\,|\,r \sim p_\text{latent}(r)\}$};

\draw [arrow, color=black] (tracinn_b.south) -- (dens.north);
\draw [arrow, color=black] ([]unfcinn_b.south) -- (jac.north);
\draw [arrow, color=black] ([]unfcinn_b.east)  -- (tracinn_b.west) node[midway,above,xshift=0.2cm]{$\{x_\text{hard}\}$};
\draw [arrow, color=black] (random.south) -- (unfcinn_b.north);
\draw [arrow, color=black] (alpha.east) -- ([yshift=-0.2cm]unfcinn_b.west);
\draw [arrow, color=black] (xr.east) -- (tracinn_b.north |- xr.east) -- (tracinn_b.north);
\draw [arrow, color=black] (alpha.east -| sigfid.north) -- (sigfid.north);
\draw [arrow, color=black] (xr.east -| sigfid.north) -- ([yshift=0.2cm]unfcinn_b.west -| sigfid.north) -- ([yshift=0.2cm]unfcinn_b.west);
\draw [arrow, color=black] ([xshift=0.2cm]dcs.north|- tracinn_b.west) -- ([xshift=0.2cm]dcs.north);
\draw [arrow, color=black] (acc.north|- tracinn_b.west) -- (accnet.north);
\draw [arrow, color=black] (accnet.south) -- (acc.north);
\draw [line, color=black] ([yshift=-1.5cm]sigfid.north|- tracinn_b.west) -- ([xshift=-0.1cm, yshift=-1.5cm]unfcinn_b.south|- tracinn_b.west);
\draw [arrow, color=black] ([xshift=0.1cm, yshift=-1.5cm]unfcinn_b.south|- tracinn_b.west) -- ([xshift=-0.2cm, yshift=-1.5cm]dcs.north|- tracinn_b.west) -- ([xshift=-0.2cm]dcs.north);

\end{tikzpicture}
\vspace{-2em}
\caption{Three-network MEM integrator evaluating
  Eq.\eqref{eq:likeli_integral} through sampling $r$. The
  Sampling-cINN is conditioned on the CP-angle $\alpha$ and the
  reco-level event $x_\text{reco}$. The Transfer network is conditioned
  on the hard-scattering event $x_\text{hard}$. For the three-network 
  setup the acceptance $\epsilon(x_\text{hard})$ is encoded in a network.}
  \label{fig:mem_setup}
\end{figure}
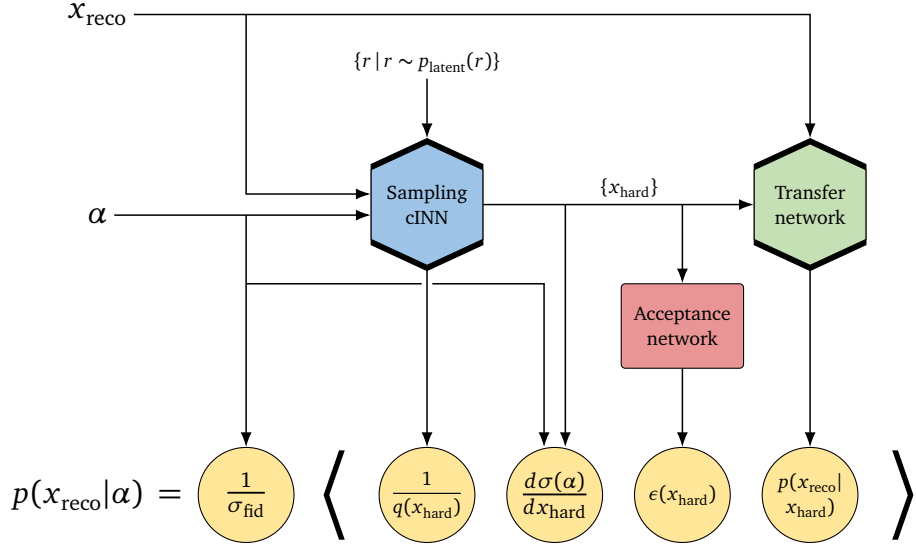

The integration in Eq.\eqref{eq:likeli_prob} is challenging, because the
differential cross section spans several orders of magnitude, and the
transfer probability typically forms a narrow peak. We solve the integral
using Monte Carlo integration sampling $x_\text{hard} \sim
q(x_\text{hard}|x_\text{reco},\alpha) \equiv q(x_\text{hard})$,
\begin{align}
  p(x_\text{reco}|\alpha)
  &= \int d x_\text{hard} \; p_\text{fid}(x_\text{hard}|\alpha) \; \pmd (x_\text{reco}|x_\text{hard})  \notag \\
  &= \XXXLangle \frac{1}{q(x_\text{hard})} \; p_\text{fid}(x_\text{hard}|\alpha) \; \pmd (x_\text{reco}|x_\text{hard})
     \XXXRangle_{x_\text{hard} \sim q(x_\text{hard})} \; ,
\end{align}
Ideally, this assumes 
\begin{align}
    \pmd (x_\text{reco}|x_\text{hard}) = p(x_\text{reco}|x_\text{hard}) \; ,
\end{align}
in which case we can use Bayes' theorem to arrive at
\begin{align}
  p(x_\text{reco}|\alpha)
  &= \XXXLangle \frac{1}{q(x_\text{hard})} \; p_\text{fid}(x_\text{hard}|\alpha) \; p(x_\text{reco}|x_\text{hard})
     \XXXRangle_{x_\text{hard} \sim q(x_\text{hard})} \notag \\
  &= \XXXLangle \frac{1}{q(x_\text{hard})} \; p(x_\text{hard}|x_\text{reco},\alpha) p(x_\text{reco}|\alpha)
     \XXXRangle_{x_\text{hard} \sim q(x_\text{hard})} \; .
\end{align}
For this integral the variance vanishes when
\begin{align}
    q(x_\text{hard})\equiv q(x_\text{hard}|x_\text{reco},\alpha) \propto p(x_\text{hard}|x_\text{reco},\alpha) \; ,
    \label{eq:ideal1}
\end{align}
where $p(x_\text{hard}|x_\text{reco},\alpha)$ corresponds to the generative unfolding probability from reco-level to parton-level~\cite{Bellagente:2020piv}.
However, in practice, we cannot expect the learned transfer probability to match
its truth counterpart perfectly. In that case the condition in
Eq.\eqref{eq:ideal1} becomes
\begin{align}
    q(x_\text{hard}|x_\text{reco},\alpha) \propto p_\text{fid}(x_\text{hard}|\alpha) \; \pmd (x_\text{reco}|x_\text{hard}) \; .
\end{align}
In both cases, we train a second conditional normalizing flow with trainable parameters $\varphi$ to encode
this optimal transformation of the integration variables,
\begin{align}
  r \sim \pl (r)
  \quad 
  \xleftrightarrow{\text{Sampling-cINN}}\quad
  x_\text{hard}(r) \sim \qmd(x_\text{hard}|x_\text{reco},\alpha) \; ,
  \label{eq:sampling_cinn}
\end{align}
which allows to parameterize the conditional sampling density as
\begin{align}
    \qmd(x_\text{hard}|x_\text{reco},\alpha)&\equiv \qmd(x_\text{hard}(r)|x_\text{reco},\alpha)=\frac{\pl(r)}{\gmd(r)}\notag\\
    &\text{with} \qquad \gmd(r)=\left\vert\frac{\partial x_\text{hard}(r; x_\text{reco},\alpha;\varphi)}{\partial r}\right\vert\;.
\end{align}
The MEM integral in Eq.\eqref{eq:likeli_3} now reads
\begin{align}
  p(x_\text{reco}|\alpha)
  &= \frac{1}{\sigma_\text{fid}(\alpha)}
  \int d r\; 
  \gmd(r) \; 
  \left[ \frac{d \sigma(\alpha)}{d x_\text{hard}} \epsmd(x_\text{hard})\,\pmd(x_\text{reco}|x_\text{hard}) \right]_{x_\text{hard}(r; x_\text{reco},\alpha;\varphi)} \notag \\
  &= \frac{1}{\sigma_\text{fid}(\alpha)}
  \left\langle
  \frac{\gmd(r)}{\pl(r)} \; 
  \left[ \frac{d \sigma(\alpha)}{d x_\text{hard}} \epsmd(x_\text{hard})\,\pmd(x_\text{reco}|x_\text{hard}) \right]_{x_\text{hard}(r; x_\text{reco},\alpha;\varphi)}
  \right\rangle_{r \sim p(r)} \; .
  \label{eq:likeli_integral}
\end{align}
The architecture of our MEM integrator is illustrated in
Fig.~\ref{fig:mem_setup}.

\section{Two-network baseline}
\label{sec:memml_improve}

In the proof-of-concept implementation of Ref.~\cite{Butter:2022vkj}
we used a series of ad-hoc fixes to stabilize the critical phase space
integration in Eq.\eqref{eq:likeli_3}. Before we present more
substantial improvements to our framework, we introduce a series of numerical improvements to our baseline two-cINN setup. For the two-network setup we assume that we can neglect the phase-space dependence of the acceptance in the MEM integration,
\begin{align}
  p(x_\text{reco}|\alpha)
  \approx \frac{1}{\sigma_\text{fid}(\alpha)} \int d x_\text{hard}\; \frac{d \sigma(\alpha)}{d x_\text{hard}} \;
  \pmd(x_\text{reco}|x_\text{hard}) \; .
  \label{eq:likeli_2} 
\end{align}
%

\subsubsection*{Single-pass integration over model parameters}

Initially, we integrate over the phase space for each theory parameter
value separately. This general approach does not make use of the fact
that the detector response does not depend on $\alpha$, and the
mapping for the importance sampling only has a small
$\alpha$-dependence.  The phase space samples $x_\text{hard} \sim
\qmd(x_\text{hard} | x_\text{reco}, \alpha)$ and the corresponding values
of $\pmd(x_\text{reco} | x_\text{hard})$ can be used to evaluate the
differential cross section for multiple points in $\alpha$.  Moreover,
parts of the cross section calculation only depend on the phase space
point and not on $\alpha$, like for example parton densities.

Consequently, we can understand the integrand for a given Monte Carlo
sample as a smooth function of $\alpha$, so the integral will also be
a smooth function of $\alpha$. This means we do not have to
fit an explicit function to the likelihood values and instead extract
a smooth log-likelihood as a function of $\alpha$.  The MEM
integration for a given $x_\text{reco}$ and a discrete set $\{\alpha \}$ can
be performed as:
\begin{enumerate}
\item For $j \in \{1, \ldots, N\}$, draw $\alpha^{(j)}$ from $\{\alpha
  \}$ randomly;
\item Using the sampling network, sample $x_\text{hard}^{(j)} \sim
  \qmd(x_\text{hard}|x_\text{reco},\alpha^{(j)})$; 
\item Evaluate the transfer probability $\pmd(x_\text{reco} |
  x_\text{hard}^{(j)})$ for each sample.
\item Evaluate the differential cross section $d \sigma(\alpha) / d
  x_\text{hard}^{(j)}$ for each sample $x_\text{hard}^{(j)}$ and
  $\alpha$;
\item Compute the MC integral Eq.\eqref{eq:likeli_2} for all
  $\alpha$ values at the same time
      \begin{align}
        p(x_\text{reco} | \alpha) &\approx
        \frac{1}{\sigma_\text{fid}(\alpha)} \frac{1}{N} \sum_{j=1}^N
        \frac{1}{\qmd(x^{(j)}_\text{hard}|x_\text{reco},\alpha^{(j)})} \; \frac{d\sigma(x_\text{hard}^{(j)}|
          \alpha)}{d x_\text{hard}} \pmd(x_\text{reco} |
        x_\text{hard}^{(j)}) 
    \end{align}
\end{enumerate}

This integral converges quickly for some events, while more statistics
are needed for others. One reason is that the peaks of the transfer
probability and the importance sampling distribution are not perfectly aligned for
some events, resulting in a higher variance. To reduce the integration
time while guaranteeing a small integration error, we compute the
integral iteratively. We specify the number of samples per iteration
as well as a minimal and maximal number of iterations. Furthermore, we
specify a threshold for the maximum relative uncertainty over the
results for all values of $\alpha$. The integration is repeated for
new batches of samples until the combined uncertainty drops below the
threshold. In practice, a batch size of 10000, at least two and at
most 15 iterations meet a target uncertainty of $2\%$. The uncertainty
on the normalized negative log-likelihood will be much smaller than
these $2\%$ because of the correlation between different $\alpha$.

\subsubsection*{Integration uncertainties}

Using this single-pass integration, the results for different $\alpha$
values become correlated, because the new algorithm ensures that the
result is a smooth function of $\alpha$.  This means that the MC
integration error cannot be easily estimated point-wise.  The
uncertainty on the likelihood ratio should be much smaller than the
uncertainty of the absolute value of the likelihood before
normalization. To account for the correlations, we use bootstrapping
to resample the integrand multiple times and propagate the resulting
replicas through the downstream tasks.  For this bootstrapping we take
our samples of the integrand $I^{(j)}(\alpha_i)$ and randomly draw $M$
batches of $N$ samples from $\{I^{(j)}(\alpha_i) \: | \: j \in \{1,
\ldots, N\}\}$ with replacement. We compute the mean over the $N$
samples per batch, defining $M$ replicas of the integral as a function
of $\alpha$. They can be used to estimate uncertainties on the
following normalized negative log-likelihoods.

Next, we can quantify the uncertainty from the training of the
transfer probability using a Bayesian
network\cite{bnn_early,bnn_early2,bnn_early3,deep_errors,Bollweg:2019skg,Kasieczka:2020vlh,Bellagente:2021yyh}. To estimate the training uncertainty we perform the phase space
integration for different samples from the distribution over the
trainable parameters. In Ref.~\cite{Butter:2022vkj} this is done by
repeating the integration for different sampled networks. However, the
idea of the single-pass integration also applies to the Bayesian
transfer probabilities. The same importance sampling distribution should
work well for different sampled networks, making the integration more
efficient. The training uncertainty estimation can be combined with
the bootstrapping procedure described above. For each replica, we do
not only resample the integrand but also compute the transfer probability
for a different sample from the distribution over the trainable
parameters.

\subsubsection*{Factorization of differential cross section}

For our example process, single-top plus Higgs production with an
anomalous CP-phase, the Lagrangian given in Eq.\eqref{eq:LttH} can be
written as
\begin{align}
    \lag = \lag_1 + \sin\alpha \; \lag_2 + \cos \alpha \: \lag_3 \; ,
\end{align}
and the squared matrix element has the corresponding form
\begin{align}
  \frac{d\sigma(x_\text{hard}| \alpha)}{d x_\text{hard}} =
  g_1
  + \sin\alpha \: g_2
  + \cos\alpha \: g_3
  + \sin\alpha \cos\alpha \: g_4
  + \sin^2 \alpha \: g_5 \; ,
\end{align}
with phase space dependent $g_i(x_\text{hard})$. This is an
example where the matrix element factorizes into an
$x_\text{hard}$-dependent and an $\alpha$-dependent part. Similar
factorization properties hold for SMEFT corrections where it is often referred to as operator morphing~\cite{Brehmer:2018eca}. For
\begin{align}
    \frac{d\sigma(x_\text{hard}| \alpha)}{d x_\text{hard}}
    = \sum_i f_i(\alpha) g_i(x_\text{hard}) 
\end{align}
the MEM integration in Eq.\eqref{eq:likeli_2} becomes
\begin{align}
  p(x_\text{reco} | \alpha)
  = \frac{1}{\sigma_\text{fid}(\alpha)}
  \sum_i f_i(\alpha) \int dx_\text{hard} \: g_i(x_\text{hard}) \; \pmd(x_\text{reco} | x_\text{hard}) \; .
\end{align}
The same can be done for the Monte Carlo estimate of the integral,
\begin{align}
    p(x_\text{reco} | \alpha) &\approx \frac{1}{\sigma_\text{fid}(\alpha)} \frac{1}{N} \sum_{j=1}^N  \frac{1}{\qmd(x^{(j)}_\text{hard}|x_\text{reco},\alpha^{(j)})} \; \frac{d\sigma(x_\text{hard}^{(j)}| \alpha)}{d x_\text{hard}} \pmd(x_\text{reco} | x_\text{hard}^{(j)}) \\
    &= \frac{1}{\sigma_\text{fid}(\alpha)} \sum_i f_i(\alpha) \frac{1}{N} \sum_{j=1}^N \:  \frac{1}{\qmd(x^{(j)}_\text{hard}|x_\text{reco},\alpha^{(j)})} \; g_i(x_\text{hard}^{(j)})\;\pmd(x_\text{reco} | x_\text{hard}^{(j)}) \; ,
\end{align}
where $x_\text{hard}^{(j)} \sim \qmd(x_\text{hard}|x_\text{reco},\alpha^{(j)})$. The exact
functional form of the integral is only preserved if the same
$x_\text{hard}^{(j)}$ are used for all values of $\alpha$.

\subsubsection*{Importance sampling trained on transfer probability}

The training of the Sampling-cINN assumes that 
the transfer network encodes $p(x_\text{reco} |
x_\text{hard})$ perfectly. The Sampling-cINN is then used for importance
sampling. From that perspective, it is less important to learn the truth
distribution
\begin{align}
    \qmd(x_\text{hard} | x_\text{reco}, \alpha)\approx p(x_\text{hard} | x_\text{reco}, \alpha) \propto p(x_\text{reco} | x_\text{hard}) p_\text{fid}(x_\text{hard} | \alpha) \; .
\end{align}
than the modeled distribution
\begin{align}
    \qmd(x_\text{hard} | x_\text{reco}, \alpha)\approx  \pmd (x_\text{reco} | x_\text{hard}) p_\text{fid}(x_\text{hard} | \alpha) \; .
\end{align}
The training data, consisting of tuples $(\alpha, x_\text{hard},
x_\text{reco})$ should then be modified by replacing the reco-level
momentum with the generated $\tilde{x}_\text{reco} \sim
\pmd(x_\text{reco} | x_\text{hard})$.   To increase the training
statistics we re-sample the reco-level momenta at the beginning of
each epoch. Because of the sharply peaked form of the transfer probability, even small deviations from the truth that do not have a significant impact on the inference performance, can lead to a significant misalignment with the importance sampling distribution. Hence, training the importance sampling on the learned transfer probability leads to a significantly better variance of the
integrations weights  and a faster convergence of the integral.

\subsubsection*{\vegas latent space refinement}

\begin{figure}[b!]
    \centering
    \includegraphics[width=0.49\textwidth]{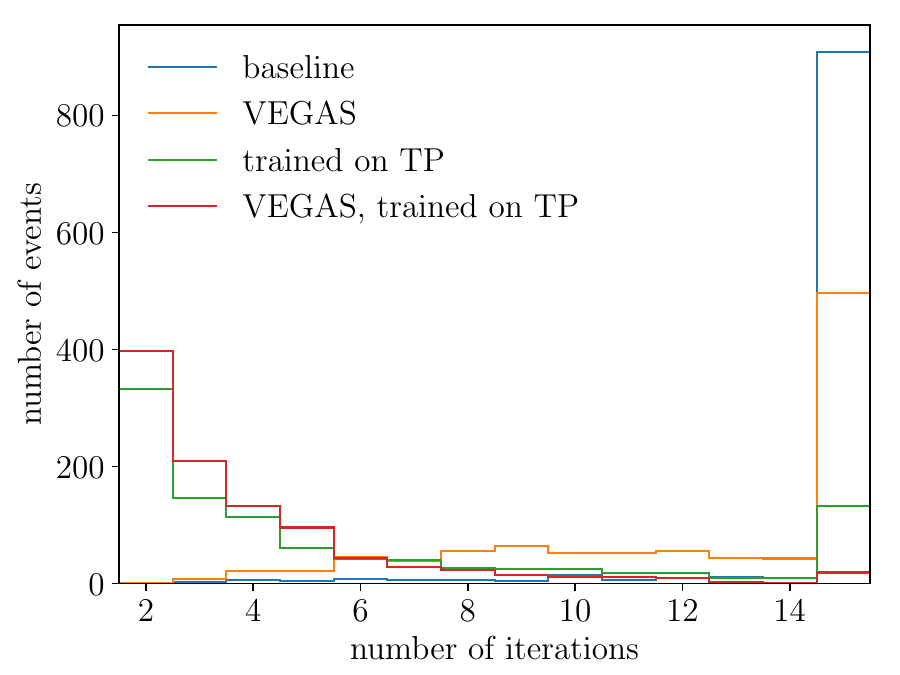}%
    \includegraphics[width=0.49\textwidth]{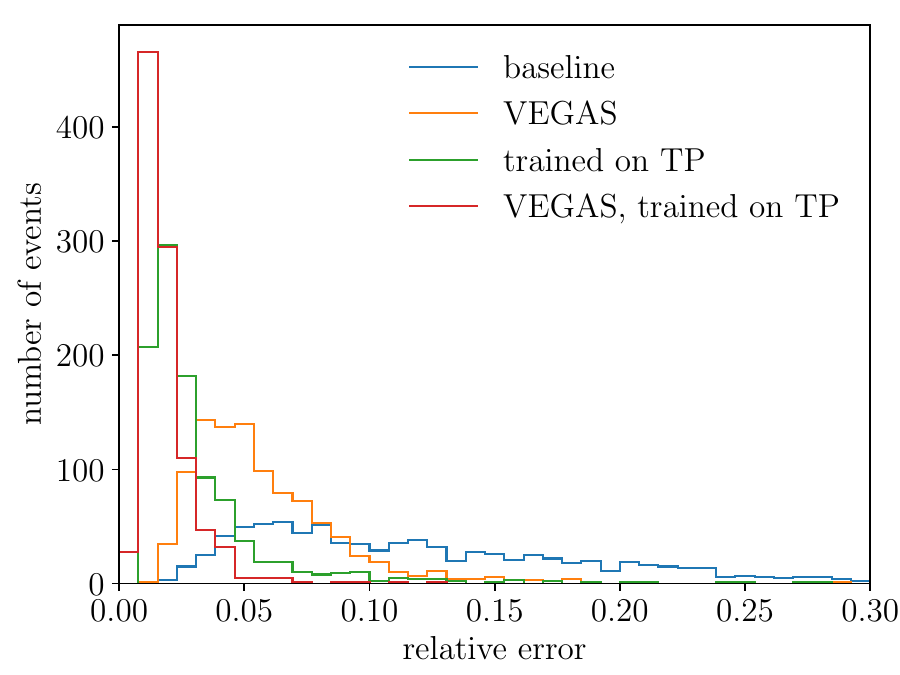}\\
    \caption{Integration performance with and without importance sampling
      trained on the transfer probability and \vegas refinement. Left:
      number of iterations (10000 samples each) to reach the 2\%
      target precision, with 2 to 15 iterations.  Right: relative
      integration error after 10 iterations of 10000 samples each.}
    \label{fig:integral}
\end{figure}

Even when the Sampling-cINN is trained on the learned transfer
probability, some events lead to a large variance in the
MEM integration. This can be solved by further adapting the proposal
distribution during the integration. Specializing the importance sampling
network for such an event is impracticable.  An alternative is to
refine the INN latent space using \vegas. Instead of directly sampling
random numbers and mapping them to phase space, we transform them with
a \vegas grid first. Note, that the grid is shared for all $\alpha$ because of the small $\alpha$ dependence of the importance sampling.
After each iteration of the integration, this
grid is adapted to reduce the variance of the integral. Because we
need to pass the integrand value back to \vegas, we choose a value in
the middle of the relevant $\alpha$-interval being evaluated. The
results from the different iterations of the integrals are combined by
weighting them by the inverse variance to reduce the overall variance
and especially the effect of early iterations where the grid is not
yet well adapted.

Figure~\ref{fig:integral} illustrates the effect of training the Sampling-cINN on the transfer probability and using \vegas refinement for the MEM integration performance with 1000 SM events and networks with a similar architecture and hyperparameters as in Ref.~\cite{Butter:2022vkj}. For our baseline, we use single-pass integration including a factorized differential cross section. While this guarantees smooth likelihood curves as a function of $\alpha$, we find that the integration uncertainty does not meet the target precision of $2\%$ within 15 iteration for most events. Running the integration with \vegas refinement improves the convergence, and the
importance sampling trained on the transfer probability leads to a even larger improvements. The combination of both methods ensures that the target precision is reached within 15 iterations for most events. 
This shows that the Sampling-cINN, trained on the transfer
function and with \vegas refinement, appears to be sufficiently 
precise to ensure fast convergence of the phase space integral. 

\subsubsection*{Two-network cINN benchmark}

\begin{figure}[b!]
    \includegraphics[width=0.33\textwidth,page=1]{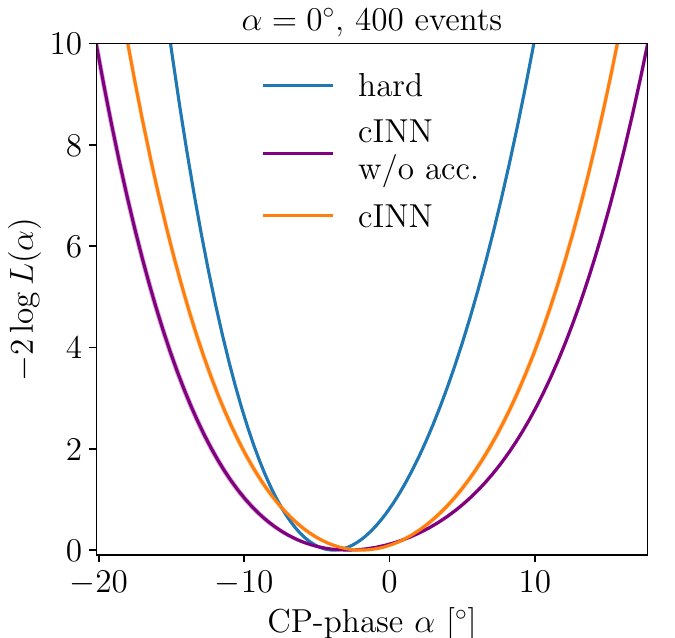}
    \includegraphics[width=0.33\textwidth,page=2]{figs/likelihoods_inn.pdf}
    \includegraphics[width=0.33\textwidth,page=3]{figs/likelihoods_inn.pdf}\\
    \includegraphics[width=0.33\textwidth,page=4]{figs/likelihoods_inn.pdf}
    \includegraphics[width=0.33\textwidth,page=5]{figs/likelihoods_inn.pdf}
    \includegraphics[width=0.33\textwidth,page=6]{figs/likelihoods_inn.pdf}\\
    \includegraphics[width=0.33\textwidth,page=1]{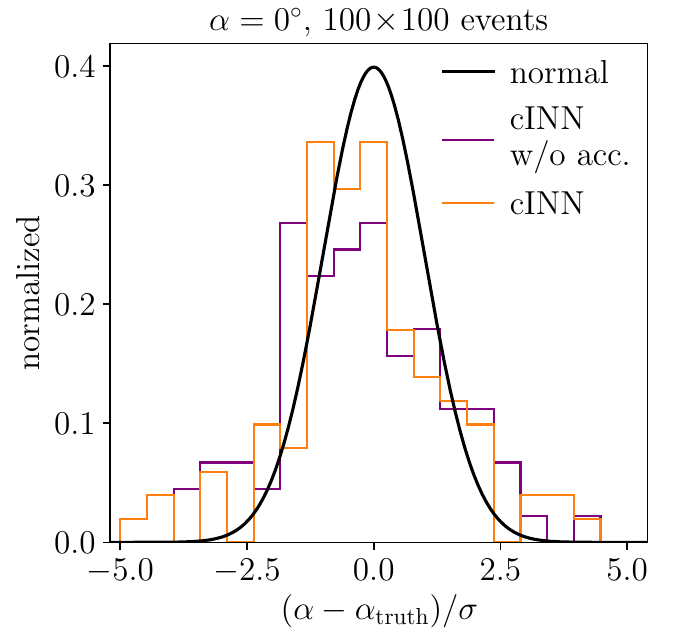}
    \includegraphics[width=0.33\textwidth,page=2]{figs/pulls_inn.pdf}
    \includegraphics[width=0.33\textwidth,page=3]{figs/pulls_inn.pdf}
    \caption{\textbf{cINN benchmark and learned acceptance:} likelihoods for different CP-angles. We use the same architecture as in
      Ref.~\cite{Butter:2022vkj}, but with the improved integration. The purple curve shows the two-network cINN benchmark and the orange curve also includes the learned acceptance. From top to bottom: likelihoods for 400 events, 10000
      events, and pulls.}
    \label{fig:inn_baseline}
\end{figure}

The purple line in Fig.~\ref{fig:inn_baseline} shows the extracted log-likelihoods
for our example process, using all improvements described in this
section, and similar architecture and hyperparameters as in
Ref.~\cite{Butter:2022vkj}.  In the top two rows we show the extracted
likelihoods from a small set of 400 events and from a large set of 10k
events. In both cases, we compare the likelihood extracted from the 
reconstructed events to the hard-process truth. Note that we show the integration uncertainties as error bands in the plots, but due to our low error threshold and the single-pass integration these are barely visible. By repeating the integration
with the same networks, we confirm that the result is perfectly stable and consistent with these uncertainties.

Performance 
issues occur when we increase the number of events. The precision of the
combined likelihood increases and leads to a systematic deviation between the hard-process and the reconstructed likelihoods. 
This is not caused by the integration, and we will 
target this shortcoming by improving the 
architecture and the training of the transfer probability.
  

\section{Acceptance classifier}
\label{sec:memml_acc}

\begin{figure}[b!]
  \includegraphics[width=0.49\textwidth,page=21]{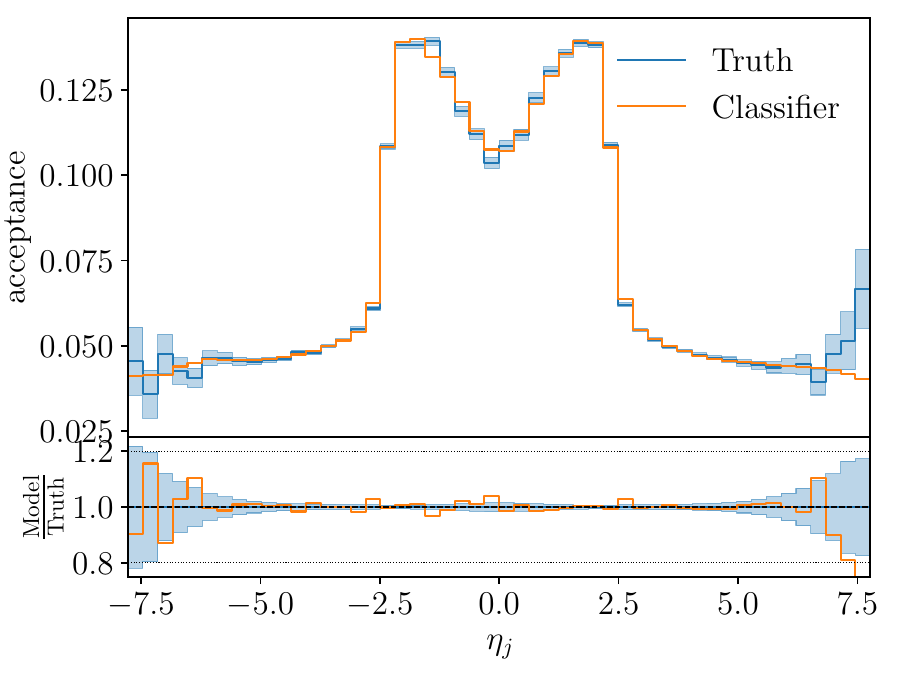}
  \includegraphics[width=0.49\textwidth,page=5]{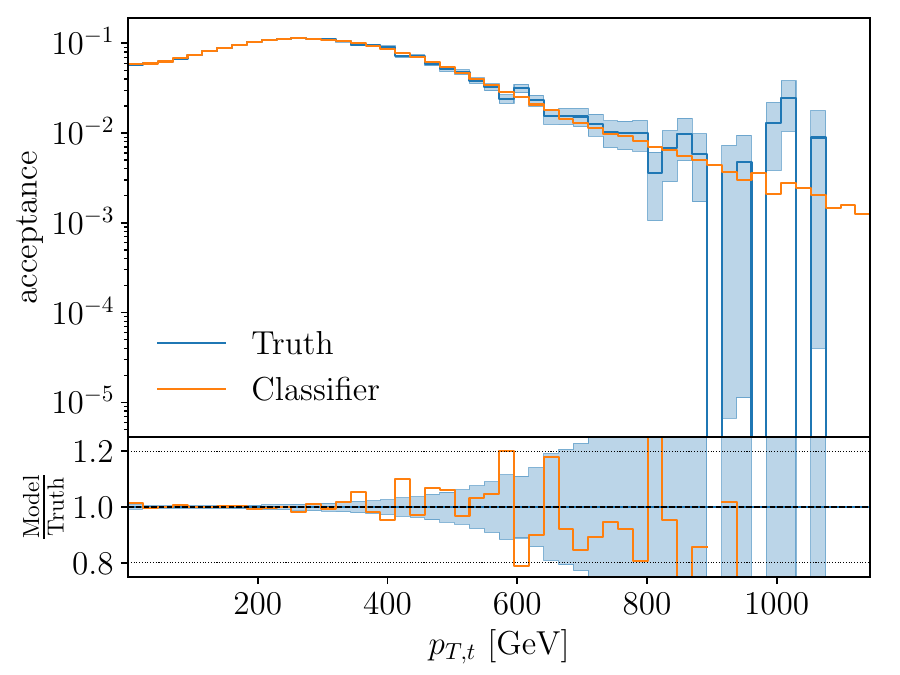}
  \caption{Truth (dashed line) and learned (solid line) acceptance as a function of different kinematic observables.}
    \label{fig:acceptance}
\end{figure}

Moving from the two-network setup in Sec.~\ref{sec:memml_improve} to
the new, three-network setup introduced in Sec.~\ref{sec:memml_setup},
we are back to the more general form of the MEM-integral,
\begin{align}
  p(x_\text{reco}|\alpha)
  = \frac{1}{\sigma_\text{fid}(\alpha)} \int d x_\text{hard}\; \frac{d \sigma(\alpha)}{d x_\text{hard}} \; \epsmd(x_\text{hard}) \;
  \pmd(x_\text{reco}|x_\text{hard}) \; .
  \label{eq:likeli_3x} 
\end{align}
The acceptance function will be encoded in a straightforward
classifier network. 
It targets the scenario where the jet from the hard process escapes detection, \ie $|\eta_j| > 2.4$, while the event is still accepted since a ISR jet is tagged instead.The two possible origins of the jets are taken into account by the transfer probability. An additional challenge is the significant drop in acceptance which is now remedied automatically by the introduction of the classifier to include the acceptance rate.

We train the classifier on a dataset of hard process configurations with
the additional information of the acceptance label. Its output then
provides $\epsmd(x_\text{hard})$ to solve Eq.\eqref{eq:likeli_3x}.
Its hyperparameters are given in
Tab.~\ref{tab:classifier_hyperparams}, and its training only takes a
few minutes. The learned and true
acceptances as a function of different kinematic observables are shown in Fig.~\ref{fig:acceptance}. Indeed,
we see a large jump in the acceptance at $|\eta_j| = 2.4$, by almost a
factor three. Also for other observables, like the $p_T$ of the top,
the acceptance varies considerably over phase space.

We then evaluate the MEM integral, now including the learned
acceptance. Comparing the new results (orange)
with the two-network baseline (purple) in Fig.~\ref{fig:inn_baseline} we
see a considerable improvement. For the small set of 400 events there is 
no bias left between the extracted likelihoods and the hard-process truth. Also 
for 10k events the large bias from Fig.~\ref{fig:inn_baseline} is
reduced to a level where it is comparable to the statistical precision. 
Even for the challenging SM-case 
$\alpha = 0^\circ$ the extracted likelihoods agrees well with the 
truth extracted from the hard process. The remaining question is how close 
we can bring the widths of the extracted likelihood-curves to the optimal 
outcome from the hard process, and if a remaining systematic bias can 
keep up with statistical improvements. From now on, we will keep the acceptance network within our MEM setup throughout the rest of our paper.

\section{Transfer diffusion}
\label{sec:diffusion}

Instead of a Transfer-cINN~\cite{Butter:2022vkj}, as discussed
in Sec.~\ref{sec:memml_setup}, we can also use other neural networks
to encode the transfer probability. The great advantages of the INN are
its stability, its controlled precision in estimating the density, and
its speed in both directions. However, these advantages come at the
prize of limited flexibility, and we can use diffusion networks to
slightly shift this balance~\cite{Butter:2023fov}. 
Conditional flow matching (CFM) networks~\cite{lipman2023flow, albergo2023building,liu2022flow} allow
for more flexibility in encoding an underlying density, with the main
disadvantage of a significant loss in speed in the likelihood
evaluation. While this speed might become a relevant factor
eventually, we compare the performance of the cINN with the CFM at face value. For a detailed introduction of conditional flow matching in the context of particle physics we refer to Ref.~\cite{Butter:2023fov} and only repeat the key points here.\medskip

The Transfer-CFM replaces the Transfer-cINN in
Eq.\eqref{eq:transfer_cinn}.  The CFM models the transformation
between a latent distribution $\pl(r)$ and a conditional phase space
distribution $\pmd(x_\text{reco}|x_\text{hard})$ inspired by a a
time-dependent process. The time evolution is described by an ordinary differential equation
\begin{align}
\frac{d x(t)}{dt} = v(x(t), t) \; ,
\label{eq:sample_ODE}
\end{align}
with the velocity field $v(x(t), t)$. The corresponding time-dependent probability
density $p(x,t)$ obeys the continuity equation
\begin{align}
\frac{\partial p(x,t)}{\partial t} + \nabla_x \left[ p(x,t) v(x,t) \right] = 0 \; .
\label{eq:continuity}
\end{align} 
%
To obtain a generative model we need a velocity field that evolves the probability density in time such that
\begin{align}
 p(x,t) \to 
 \begin{cases}
  \pmd(x) \approx \pd(x) \qqquad & t \to 0 \\
  \pl(x) = \normal(x;0,1) \qqquad & t \to 1  \; .
\end{cases} 
\label{eq:fm_limits}
\end{align}
To construct this velocity field we start from a sample-conditional diffusion trajectory
\begin{align}
    x(t|x_0) 
    = (1-t) x_0 + t r
\to \begin{cases}
      x_0 \qqquad & t \to 0 \\
      r \sim \normal(0,1) \qqquad & t \to 1 \; ,
\end{cases} 
\label{eq:gaussian_probability_path_reparametrization}
\end{align}
that evolves the phase space sample $x_0$ towards a latent space sample. The associated sample-conditional velocity field directly follows from the ODE Eq.\eqref{eq:sample_ODE}
\begin{align}
   v(x(t|x_0),t| x_0) 
   &= \frac{d}{dt} \left[ (1-t) x_0 + t r \right] = - x_0 + r \; .
\label{eq:conditional_velocity}
\end{align}
The desired velocity field for the generative model is then given by~\cite{lipman2023flow} 
\begin{align}
    v(x,t) = \int dx_0 \; \frac{v(x,t|x_0)p(x,t|x_0)\pd(x_0)}{p(x,t)} \; .
    \label{eq:velocity}
\end{align}
Learning the velocity field from data is a straightforward regression task and can again be reformulated in terms of the conditional velocity field~\cite{lipman2023flow}   
%
\begin{align}
    \loss_\text{FM} 
    &= \XLangle \left[ v_\theta(x,t) - v(x,t) \right]^2 \XRangle_{t, x\sim p(x,t)} \; , \notag \\
    & \bigg\downarrow \;\text{reparametrization + neglecting constants}  \notag\\
    \loss_\text{CFM} &= \XLangle \left[ v_\theta(x(t|x_0),t) - v(x(t|x_0),t|x_0) \right]^2 \XRangle_{t\sim U(0,1),x_0\sim \pd, r\sim \normal(0,1)}
\label{eq:FMloss}
\end{align}
%
Once the model is trained to encode the velocity it defines a bijective mapping between the latent and the phase space via numerically solving the ODE Eq.\eqref{eq:sample_ODE}. Crucially for our application
the Jacobian of this transformation 
is tractable through another ODE~\cite{chen2019neural}
\begin{align}
\frac{d \log p(x(t), t)}{dt} = - \nabla_x v(x(t),t).
\label{eq:logp_ODE}
\end{align}
To calculate the likelihood of a phase space sample $x$ we map it to the latent space according to Eq.\eqref{eq:sample_ODE} and calculate the jacobian determinant of this transformation according to Eq.\eqref{eq:logp_ODE} 
\begin{align}
r(x) &= x + \int_0^1 v_\theta (x,t) dt \;  \quad \text{with} \quad
\left|\frac{\partial r}{\partial x} \right| 
     = \exp \left( \int_0^1 dt \nabla_x v_\theta(x(t),t) \right) \; . \\
     &\Rightarrow \qquad p(x) = \pl (r(x))\exp \left( \int_0^1 dt \nabla_x v_\theta(x(t),t) \right) 
\label{eq:cfm_likelihood}
\end{align}
Solving the ODEs numerically with the required precision takes $\mathcal{O}(100)$ evaluations of the function. For the transformation ODE this is relatively fast as the function is just the velocity, i.e. the neural network. For the likelihood ODE however evaluating the function means calculating the gradients of all components of the velocity with respect to the inputs, making likelihood calculation significantly slower.

The
hyperparameters of our CFM network are given in
Tab.~\ref{tab:cfm_transfermer_hyperparams}. It is 
straightforward to replace the Transfer-cINN with a Transfer-CFM 
in our MEM architecture, so we can benchmark the performance gain through the increased expressivity, at the possible expense of speed.

\begin{figure}[t]
    \includegraphics[width=0.33\textwidth,page=1]{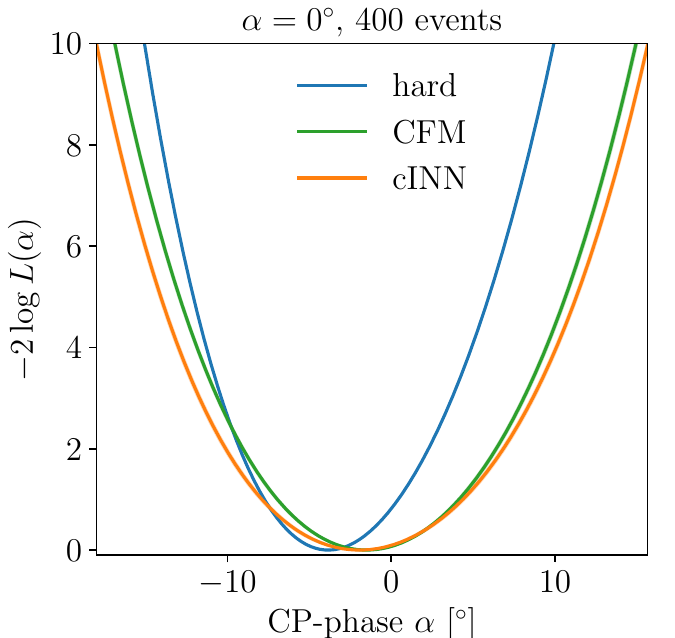}
    \includegraphics[width=0.33\textwidth,page=2]{figs/likelihoods_cfm.pdf}
    \includegraphics[width=0.33\textwidth,page=3]{figs/likelihoods_cfm.pdf}\\
    \includegraphics[width=0.33\textwidth,page=4]{figs/likelihoods_cfm.pdf}
    \includegraphics[width=0.33\textwidth,page=5]{figs/likelihoods_cfm.pdf}
    \includegraphics[width=0.33\textwidth,page=6]{figs/likelihoods_cfm.pdf}\\
    \includegraphics[width=0.33\textwidth,page=1]{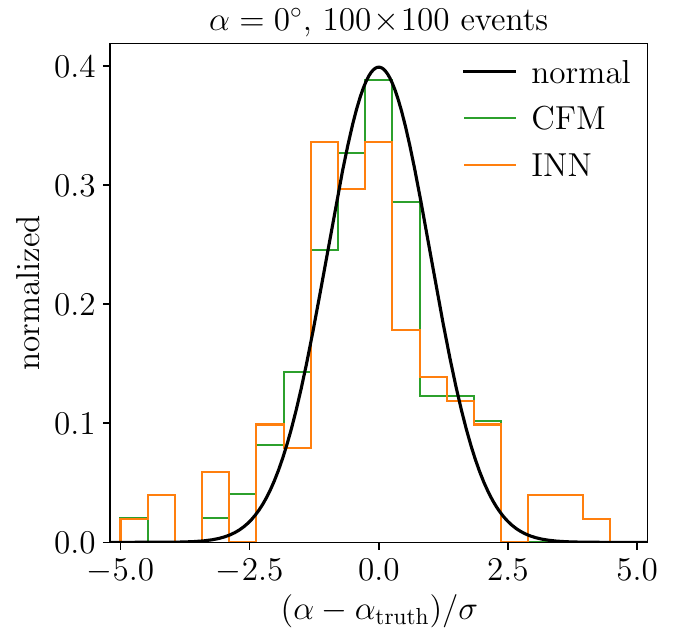}
    \includegraphics[width=0.33\textwidth,page=2]{figs/pulls_cfm.pdf}
    \includegraphics[width=0.33\textwidth,page=3]{figs/pulls_cfm.pdf}
    \caption{\textbf{Transfer-CFM:} likelihoods for different CP-angles.
      We compare the cINN baseline
      with a CFM diffusion network, both including the learned acceptance. From top to bottom: likelihoods for 400
      events, 10000 events, and pulls.}
    \label{fig:cfm_acceptance}
\end{figure}

The likelihoods extracted with the help of the Transfer-CFM are illustrated in Fig.~\ref{fig:cfm_acceptance}
and can be compared to the same MEM setup, but with a Transfer-cINN in Fig.~\ref{fig:inn_baseline}.
For 400 events the difference between the Transfer-cINN and the Transfer-CFM is not visible,
suggesting that both of them work extremely well given the statistical limitations and the
phase space integration. There is no systematic bias, and the width of the extracted likelihoods 
are close to the optimal hard-process curves.

For the high-precision case with 10k events the Transfer-CFM leads to a significant improvement
over the cINN architecture.
Now, the picture is the same as for 400 events, where the extracted likelihoods do not show any 
significant bias, and the extracted likelihoods are extremely close to the optimal information.

\section{Combinatorics transformer}
\label{sec:tramsformer}

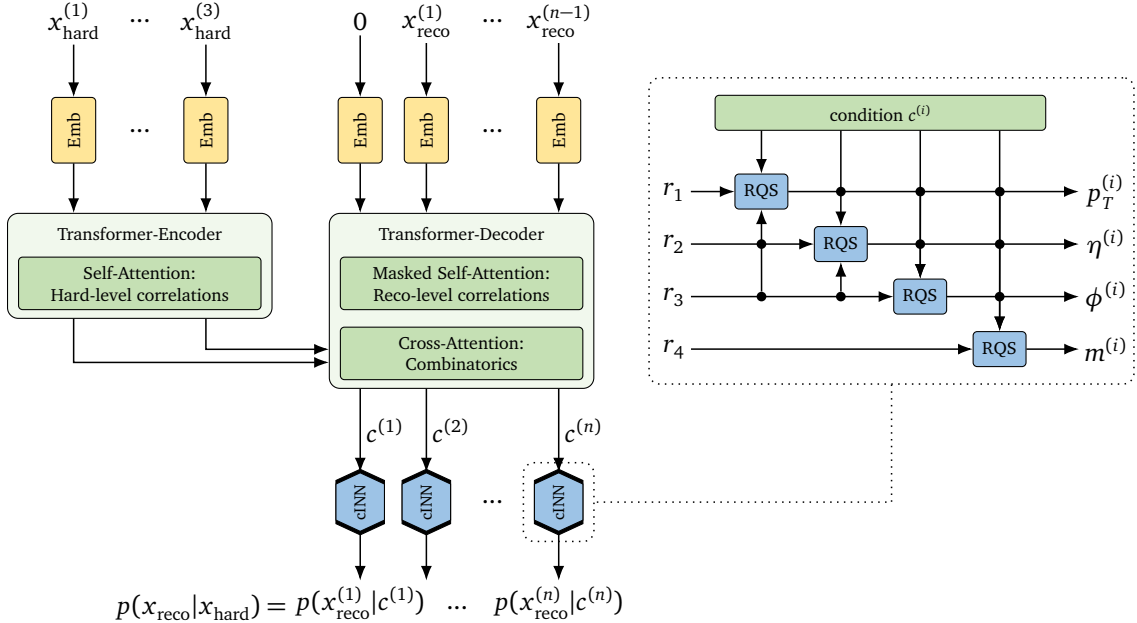
\begin{figure}[b!]
    \definecolor{Rcolor}{HTML}{E99595}
\definecolor{Gcolor}{HTML}{C5E0B4}
\definecolor{Gcolorlight}{HTML}{F1F8ED}
\definecolor{Bcolor}{HTML}{9DC3E6}
\definecolor{Ycolor}{HTML}{FFE699}
\definecolor{Ycolor_light}{HTML}{FFF7DE}

\tikzstyle{expr} = [rectangle, rounded corners=0.3ex, minimum width=1.5cm, minimum height=1cm, text centered, align=center, inner sep=0, fill=Ycolor, font=\large, draw]

\tikzstyle{small_cinn} = [double arrow, double arrow head extend=0cm, double arrow tip angle=130, inner sep=0, align=center, minimum width=1.1cm, minimum height=0.5cm, fill=Bcolor, draw]

\tikzstyle{small_cinn_black} = [small_cinn, minimum height=1.5cm, fill=black]

\tikzstyle{transformer} = [rectangle, rounded corners, minimum width=6cm, minimum height=2.4cm, font=\large, fill=Gcolorlight, draw]

\tikzstyle{attention} = [rectangle, rounded corners=0.3ex, minimum width=5.5cm, minimum height=1.2cm, align=center, fill=Gcolor, draw, font=\large]

\tikzstyle{txt_huge} = [align=center, font=\Huge, scale=2]
\tikzstyle{txt} = [align=center, font=\LARGE]

\tikzstyle{arrow} = [thick,-{Latex[scale=1.0]}, line width=0.2mm, color=black]
\tikzstyle{line} = [thick, line width=0.2mm, color=black]

\tikzstyle{box} = [rectangle, rounded corners=2ex, minimum width=1.4cm, minimum height=6cm, draw=white, line width=0.6mm]
\tikzstyle{upperbox} = [rectangle, rounded corners, minimum width=10cm, minimum height=1cm, align=center, draw=Bcolor, line width=0.6mm, font=\LARGE]
\tikzstyle{dot} = [circle, fill, minimum width= 2.2mm, inner sep=0pt]
\tikzstyle{rqs} = [rectangle, rounded corners=0.3ex, minimum width=1.2cm, minimum height=0.8cm, text centered, align=center, inner sep=0, fill=Bcolor, font=\large, draw]

\tikzstyle{border} = [rectangle, minimum width=11cm, minimum height=7.0cm, draw, dotted, line width=0.2mm, rounded corners=0.8ex]
\tikzstyle{smallborder} = [rectangle, minimum width=1.6cm, minimum height=1.8cm, draw, dotted, line width=0.2mm, rounded corners=0.8ex]
\tikzstyle{dottedline} = [thick, line width=0.2mm, color=black, dotted]

\begin{tikzpicture}[node distance=2cm, scale=0.58, every node/.style={transform shape}]

\node (hard1) [txt] {$x_{\text{hard}}^{(1)}$};
\node (hard2) [txt, right of=hard1, xshift=-0.5cm] {$...$};
\node (hard3) [txt, right of=hard2, xshift=-0.5cm] {$x_{\text{hard}}^{(3)}$};

\node (emb_hard1) [expr, below of=hard1, yshift=-0.5cm, rotate=90]{Emb};
\node (emb_hard2) [txt, below of=hard2, yshift=-0.5cm] {$...$};
\node (emb_hard3) [expr, below of=hard3,yshift=-0.5cm,  rotate=90]{Emb};

\node (TE) [transformer, below of=emb_hard2, yshift=-1.1cm, text width=4cm,
text depth=1.5cm, align=center] {Transformer-Encoder};
\node (TE_att) [attention, below of=TE, yshift=1.6cm] {Self-Attention: \\
Hard-level correlations};

\node (reco1) [txt, right of=hard3, xshift=1.5cm] {$0$};
\node (reco2) [txt, right of=reco1, xshift=-0.5cm] {$x_{\text{reco}}^{(1)}$};
\node (reco3) [txt, right of=reco2, xshift=-0.5cm] {$...$};
\node (reco6) [txt, right of=reco3, xshift=-0.5cm] {$x_{\text{reco}}^{(n-1)}$};

\node (emb_reco1) [expr, below of=reco1, yshift=-0.5cm, rotate=90]{Emb};
\node (emb_reco2) [expr, below of=reco2, yshift=-0.5cm, rotate=90]{Emb};
\node (emb_reco3) [txt, below of=reco3, yshift=-0.5cm] {$...$};
\node (emb_reco6) [expr, below of=reco6, yshift=-0.5cm, rotate=90]{Emb};

\node (TD) [transformer, right of=TE, xshift=5.3cm, yshift=-0.8cm, text width=4cm,
text depth=3.1cm, align=center, minimum height=4cm] {Transformer-Decoder};
\node (TD_att) [attention, right of=TE_att, xshift=5.3cm] {Masked Self-Attention: \\
Reco-level correlations};
\node (TD_crossatt) [attention, below of=TD_att, yshift=0.4cm] {Cross-Attention: \\
Combinatorics};

\node (inn1_b) [small_cinn_black, below of=reco1, yshift=-9cm, rotate=90]{cINN};
\node (inn2_b) [small_cinn_black, below of=reco2, yshift=-9cm, rotate=90]{cINN};
\node (inn6_b) [small_cinn_black, below of=reco6, yshift=-9cm, rotate=90]{cINN};
\node (inn1) [small_cinn, below of=reco1, yshift=-9cm, rotate=90]{cINN};
\node (inn2) [small_cinn, below of=reco2, yshift=-9cm, rotate=90]{cINN};
\node (inn3) [txt, below of=reco3, yshift=-9cm]{$...$};
\node (inn6) [small_cinn, below of=reco6, yshift=-9cm, rotate=90]{cINN};

\node (prob1) [txt, below of=inn1, yshift=-0.3cm]{$p(x_\text{reco}^{(1)} | c^{(1)})$};
\node (prob2) [txt, below of=inn2, yshift=-0.3cm, xshift=0.7cm, yshift=-0.1cm]{$...$};
\node (prob6) [txt, below of=inn6, yshift=-0.3cm]{$p(x_\text{reco}^{(n)} | c^{(n)})$};
\node (prob) [txt, left of=prob1, yshift=-0.1cm, xshift=-1.6cm]{$p(x_\text{reco} | x_\text{hard}) = $};

\draw [arrow, color=black] (hard1.south) -- (emb_hard1.east);
\draw [arrow, color=black] (hard3.south) -- (emb_hard3.east);

\draw [arrow, color=black] (emb_hard1.west) -- (TE.north -| emb_hard1.west);
\draw [arrow, color=black] (emb_hard3.west) -- (TE.north -| emb_hard3.west);

\draw [arrow, color=black] (TE.south -| emb_hard1.west) --  ([yshift=-1cm]TE.south -| emb_hard1.west) -- ([yshift=-1cm]TE.south -| TD.west) ; 
\draw [arrow, color=black] (TE.south -| emb_hard3.west) --  ([yshift=-0.7cm]TE.south -| emb_hard3.west) -- ([yshift=-0.7cm]TE.south -| TD.west);

\draw [arrow, color=black] (reco1.south) -- (emb_reco1.east);
\draw [arrow, color=black] (reco2.south) -- (emb_reco2.east);
\draw [arrow, color=black] (reco6.south) -- (emb_reco6.east);

\draw [arrow, color=black] (emb_reco1.west) -- (TD.north -| emb_reco1.west);
\draw [arrow, color=black] (emb_reco2.west) -- (TD.north -| emb_reco2.west);
\draw [arrow, color=black] (emb_reco6.west) -- (TD.north -| emb_reco6.west);

(A) (B);

\draw [arrow, color=black] (TD.south -| emb_reco1.west)  -- node [text width=1.5cm,midway, font=\LARGE, right] {$c^{(1)}$} (inn1.east -| emb_reco1.west);
\draw [arrow, color=black] (TD.south -| emb_reco2.west)  -- node [text width=1.5cm,midway, font=\LARGE, right] {$c^{(2)}$} (inn2.east -| emb_reco2.west);
\draw [arrow, color=black] (TD.south -| emb_reco6.west)  -- node [text width=1.5cm,midway, font=\LARGE, right] {$c^{(n)}$}  (inn6.east -| emb_reco6.west);

\draw [arrow, color=black] (inn1.west -| emb_reco1.west) -- (prob1.north -| emb_reco1.west);
\draw [arrow, color=black] (inn2.west -| emb_reco2.west) -- (prob6.north -| emb_reco2.west);
\draw [arrow, color=black] (inn6.west -| emb_reco6.west) -- (prob6.north -| emb_reco6.west);
\node (z1) [txt, right of=TD, yshift=2.5cm, xshift=2.8cm] {$r_1$};
\node (z2) [txt, below of=z1, yshift=0.8cm] {$r_2$};
\node (z3) [txt, below of=z2, yshift=0.8cm] {$r_3$};
\node (z4) [txt, below of=z3, yshift=0.8cm] {$r_4$};
\node (px) [txt, right of=z1, xshift=7.8cm] {$p_T^{(i)}$};
\node (py) [txt, below of=px, yshift=0.8cm] {$\eta^{(i)}$};
\node (pz) [txt, below of=py, yshift=0.8cm] {$\phi^{(i)}$};
\node (m) [txt, below of=pz, yshift=0.8cm] {$m^{(i)}$};
\node (RQS1) [rqs, right of=z1, xshift=0.0cm]{RQS};
\node (RQS2) [rqs, right of=z2, xshift=1.8cm]{RQS};
\node (RQS3) [rqs, right of=z3, xshift=3.6cm]{RQS};
\node (RQS4) [rqs, right of=z4, xshift=5.4cm]{RQS};
\node (c1r2) [dot,below of=RQS1, yshift=0.8cm]{};
\node (c1r3) [dot,below of=c1r2, yshift=0.8cm]{};
\node (c2r1) [dot,above of=RQS2, yshift=-0.8cm]{};
\node (c2r3) [dot,below of=RQS2, yshift=0.8cm]{};
\node (c3r2) [dot,above of=RQS3, yshift=-0.8cm]{};
\node (c3r1) [dot,above of=c3r2, yshift=-0.8cm]{};
\node (c4r3) [dot,above of=RQS4, yshift=-0.8cm]{};
\node (c4r2) [dot,above of=c4r3, yshift=-0.8cm]{};
\node (c4r1) [dot,above of=c4r2, yshift=-0.8cm]{};
\draw [arrow, color=black] ([xshift=0.4cm]z1.center) -- (RQS1.west);
\draw [arrow, color=black] ([xshift=0.4cm]z2.center) -- (RQS2.west);
\draw [arrow, color=black] ([xshift=0.4cm]z3.center) -- (RQS3.west);
\draw [arrow, color=black] ([xshift=0.4cm]z4.center) -- (RQS4.west);
\draw [arrow, color=black] (RQS1.east) -- ([xshift=-0.6cm]px.center);
\draw [arrow, color=black] (RQS2.east) -- ([xshift=-0.6cm]py.center);
\draw [arrow, color=black] (RQS3.east) -- ([xshift=-0.6cm]pz.center);
\draw [arrow, color=black] (RQS4.east) -- ([xshift=-0.6cm]m.center);
\draw [arrow, color=black] (c1r3.north) -- (RQS1.south);
\draw [arrow, color=black] (c2r3.north) -- (RQS2.south);
\draw [arrow, color=black] (c3r1.south) -- (RQS3.north);
\draw [arrow, color=black] (c4r1.south) -- (RQS4.north);
\node (condition)[transformer, above of=z1, yshift=-0.2cm, xshift=4.7cm, minimum width=7.5cm, minimum height=0.8cm, fill=Gcolor, rounded corners=0.3ex]{condition $c^{(i)}$};
\draw [arrow, color=black] (condition.south -| RQS1.center) -- (RQS1.north);
\draw [arrow, color=black] (condition.south -| RQS2.center) -- (RQS2.north);
\draw [arrow, color=black] (condition.south -| RQS3.center) -- (RQS3.north);
\draw [arrow, color=black] (condition.south -| RQS4.center) -- (RQS4.north);

\node (border) [border, below of=condition, xshift=0.25cm, yshift=-0.7cm]{};
\node (cinnborder) [smallborder, below of=reco6, yshift=-9.0cm]{};
\draw [dottedline] (border.south) -- (border.south |- cinnborder.east) -- (cinnborder.east);
\end{tikzpicture}
    \caption{Left: transformer combined with cINN, encoding the transfer probability. Right: cINN used to learn individual momenta, where $r$ is the usual latent space to parametrize a generative model.}
    \label{fig:schema_transfermer}
\end{figure}

In our last step, we introduce a transformer~\cite{Mikuni:2020wpr,Finke:2023veq,Butter:2023fov} to combine the
stability and precision of the Transfer-cINN and Transfer-CFM with an appropriate
treatment of jet combinatorics~\cite{Dillon:2021gag}. 
The structure follows the idea that the transfer probability turns a sequence of
parton-level momenta into a sequence of reco-level momenta. The Transfer-Transformer, in short Transfermer, 
should be ideal to encode the correlations between the different particles, without
relying on locality or any other physics-inspired requirement.

\subsubsection*{Transfermer}

\begin{figure}[t]
    \includegraphics[width=0.33\textwidth,page=1]{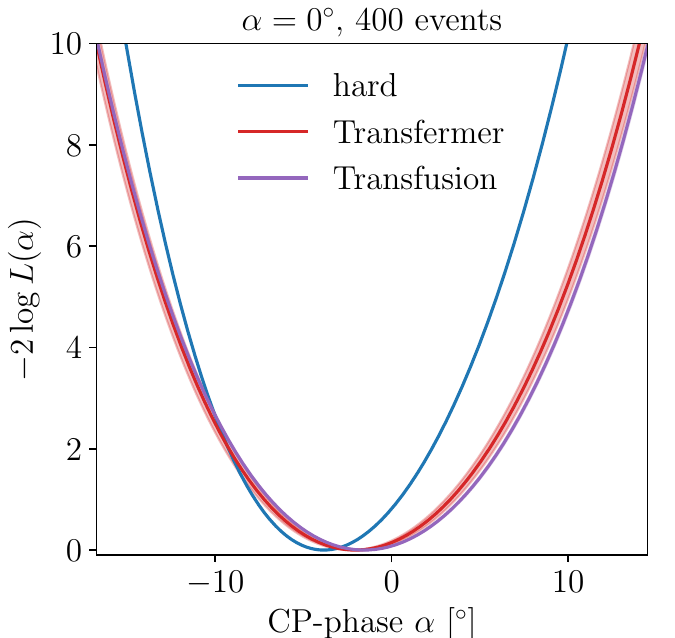}
    \includegraphics[width=0.33\textwidth,page=2]{figs/likelihoods_trans.pdf}
    \includegraphics[width=0.33\textwidth,page=3]{figs/likelihoods_trans.pdf}\\
    \includegraphics[width=0.33\textwidth,page=4]{figs/likelihoods_trans.pdf}
    \includegraphics[width=0.33\textwidth,page=5]{figs/likelihoods_trans.pdf}
    \includegraphics[width=0.33\textwidth,page=6]{figs/likelihoods_trans.pdf}\\
    \includegraphics[width=0.33\textwidth,page=1]{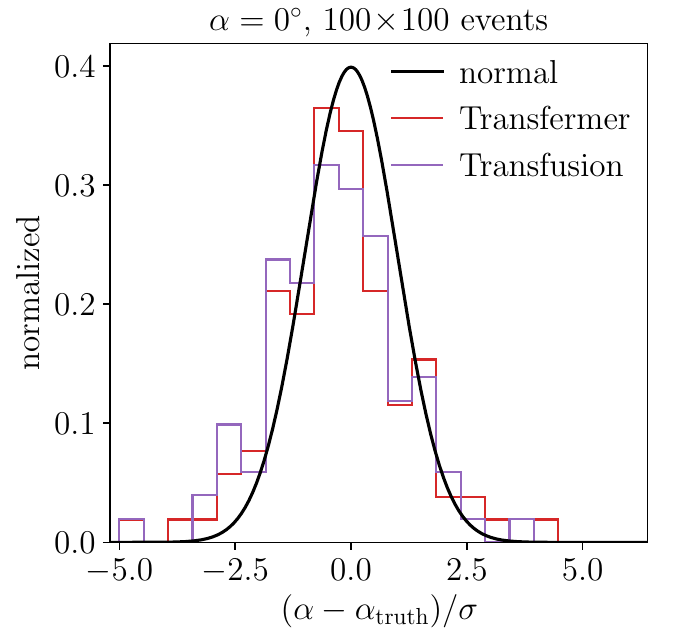}
    \includegraphics[width=0.33\textwidth,page=2]{figs/pulls_trans.pdf}
    \includegraphics[width=0.33\textwidth,page=3]{figs/pulls_trans.pdf}
    \caption{\textbf{Transfermer and Transfusion:} likelihoods for different CP-angles using a transformer for the transfer
      probability, combined with a cINN or a CFM network, respectively. From top to bottom: likelihoods for 400
      events, 10000 events, and pulls. Only the Transfermer curve includes the training uncertainties estimated with the Bayesian network.}
    \label{fig:transfermer_fixed}
\end{figure}

The challenge of using a transformer in our MEM setup is
that it is not invertible and does not guarantee a tractable
Jacobian. We can solve this problem by making the architecture
autoregressive at the level of reco-level momenta and 
splitting it into two parts, as illustrated in 
the left panel of Fig.~\ref{fig:schema_transfermer}: (i) the 
transformer encodes the correlations between the parton-level and reco-level 
objects. Their cross-correlation describes the input-output combinatorics; (ii) a small
and universal cINN encodes the correlations between the momentum components
of a single particle, conditioned on the output of the
transformer $c^{(i)}$. 

To guarantee a tractable Jacobian of the full normalizing flow, we apply
an autoregressive factorization of the transfer probability defined in 
Eq.\eqref{eq:varjet_transfer},
\begin{align}
    p(x_\text{reco}|x_\text{hard}) = \prod_{i=1}^n
    p(x_\text{reco}^{(i)} | c(e_\text{reco}^{(0)}, \ldots, e_\text{reco}^{(i-1)}, e_\text{hard})) \; .
    \label{eq:transfermer_prob}
\end{align}
The function $c$ denotes the transformer encoding. We define a special
starting token $e_\text{reco}^{(0)}$, shift the inputs by one and mask
the self-attention matrix using a triangular mask to ensure that every
momentum is only conditioned on the previous
momenta. $e^{(i)}_\text{reco}$ and $e^{(i)}_\text{hard}$ denote the
particle-wise embeddings of the momenta and their position. We define 
this embedding as the
concatenation of the momenta and their one-hot-encoded position in the
event, padded with zeros. Using a single linear layer instead
of the zero-padding does not lead to any performance
improvements. 
We then sample from the transfer probability iteratively, which  requires $n$ Transfermer evaluations,
\begin{align}
    p(x_\text{reco}^{(i)} | x_\text{hard})
    \equiv p(x_\text{reco}^{(i)} | c(e_\text{reco}^{(0)}, \ldots, e_\text{reco}^{(i-1)}, e_\text{hard})) \; .
    \label{eq:transfermer_sampling}
\end{align}
Since all $c^{(i)}$ can be
computed in a single step from the reco-level momenta, density
estimation and training this model is very fast. This is also the way
the Transfermer is used during the MEM integration. 

The transfer probability in Eq.\eqref{eq:transfermer_prob} still
has to be converted into a probability distribution for the 4-momentum
components of the external particles. To encode massless and massive 
particles in the same cINN we factorize it into
\begin{align}
    p(x_\text{reco}^{(i)}|c^{(i)}) = 
    p(p_T^{(i)}, \eta^{(i)}, \phi^{(i)}|c^{(i)}) \times
    p(m^{(i)}|p_T^{(i)}, \eta^{(i)}, \phi^{(i)}, c^{(i)}) \; ,
\end{align}
such that the generation of the mass direction can be omitted without
affecting the other three components. The corresponding cINN architecture is 
given in the right panel of Fig.~\ref{fig:schema_transfermer}.
Rational quadratic spline coupling layers 
model the one-dimensional distributions. By transforming each momentum component once and
conditioning it on the other components and the 
transformer output, using a feed-forward network, we build a
minimal cINN that is able to model the correlations between the
momentum components. 

In practice, we use
normalized versions of $\log p_T$ and $\log m$ as inputs for the
network and map them to Gaussian latent spaces. Similarly, we map $\phi$ and
$\eta$ to uniform latent spaces, taking into account the
detector-level $\eta$ cuts. For $\phi$ we use periodic RQS splines~\cite{Ackerschott:2023nax}.
The cINN for single
momenta and the transformer are trained jointly by minimizing the
negative log-likelihood loss $\loss = - \log \pmd(x_\text{reco}|x_\text{hard})$.

We implement the Transfermer with the standard \pytorch~\cite{pytorch} transformer module and the cINN architecture described above. The hyperparameters are given in Tab.~\ref{tab:cfm_transfermer_hyperparams}. 
In Fig.~\ref{fig:transfermer_fixed}, we show the likelihoods for the Transfermer architecture. This plot shows much larger error bands because they also include the systematic uncertainty from the Transfermer training, estimated with a Bayesian network. For the other architectures, we omit these due to runtime constraints. The likelihoods can be compared to the cINN results in Fig.~\ref{fig:inn_baseline}, and we see that their bias and accuracy have improved. Even for 10k events, the likelihoods are 
largely unbiased, albeit not significantly better than for the 
Transfer-CFM from Fig.~\ref{fig:cfm_acceptance}.
The Transfermer architecture can be easily generalized to support variable numbers of reco-level jets. We show this extension in Appendix~\ref{sec:varjet_perminv} but do not find any additional improvements for our reference process. Furthermore, we show how sensitive this architecture is to the choice of simulation tool in Appendix~\ref{sec:herwig}.

\subsubsection*{Transfusion}

As a last transfer architecture we consider the CFM equivalent of the Transfermer, an autoregressive Transfusion. We keep the autoregressive structure and the masked self-attention from Fig.~\ref{fig:schema_transfermer} and simply replace the small cINN with a small CFM network to generate the individual particle momenta. The CFM learning task is a simple regression of the velocity field. As long as we can track gradients through the network, we obtain a tractable Jacobian according to Eq.\eqref{eq:logp_ODE}. The velocity of the $i^\text{th}$ particle is then denoted in analogy to Eq.\eqref{eq:transfermer_prob} 
\begin{align}
    v^{(i)}(x_\text{reco}^{(i)}(t), t| c(e_\text{reco}^{(0)}, \ldots, e_\text{reco}^{(i-1)}, e_\text{hard})) \; ,
    \label{eq:transfusionAR_velocity}
\end{align}
From the velocity field the likelihoods are again obtained by solving the ODEs Eq.\eqref{eq:cfm_likelihood}, in the Transfusion setup now autoregressively for each particle.
For on-shell and off-shell particles, we use two different small CFMs, one 3-dimensional and one 4-dimensional. This setup outperforms just using the same 4-dimensional network and discarding the generated masses for on-shell particles. The hyperparameters of the Transfusion network are given in Tab.~\ref{tab:transfusion_hyperparams}. 

We show the MEM likelihoods obtained with the Transfusion in Fig.~\ref{fig:transfermer_fixed} and find that they are indistinguishable from the Transfermer results. This indicates that the difference between the cINN and CFM likelihoods can be attributed to cINN issues with the jet combinatorics. Outsourcing this task to the transformer significantly improves the performance. For the CFM the corresponding improvement is minimal.

\section{Outlook}
\label{sec:outlook}

\begin{figure}[t!]
    \includegraphics[width=0.49\textwidth,page=23]{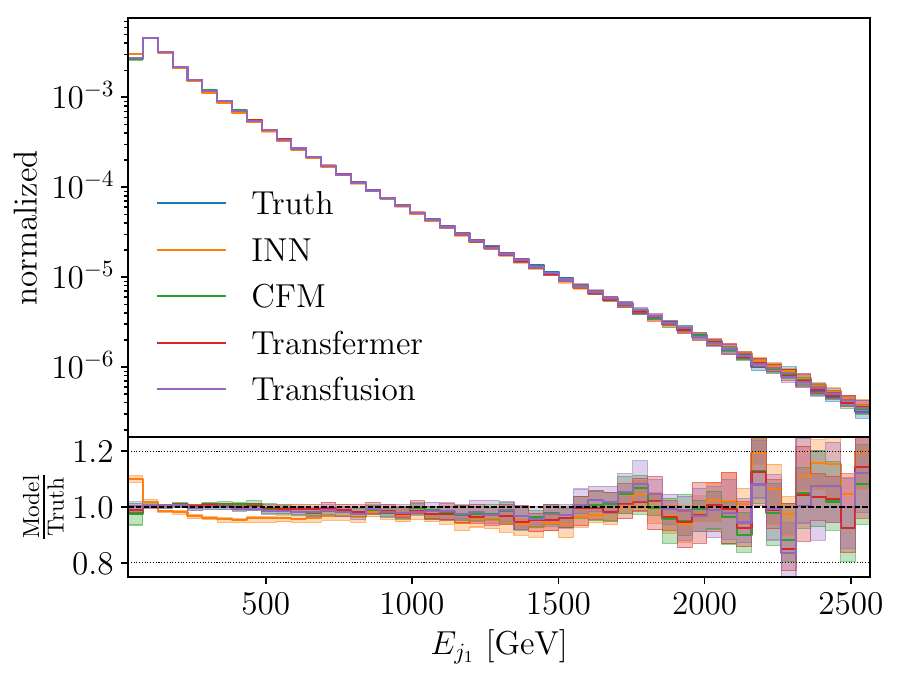}
    \includegraphics[width=0.49\textwidth,page=27]{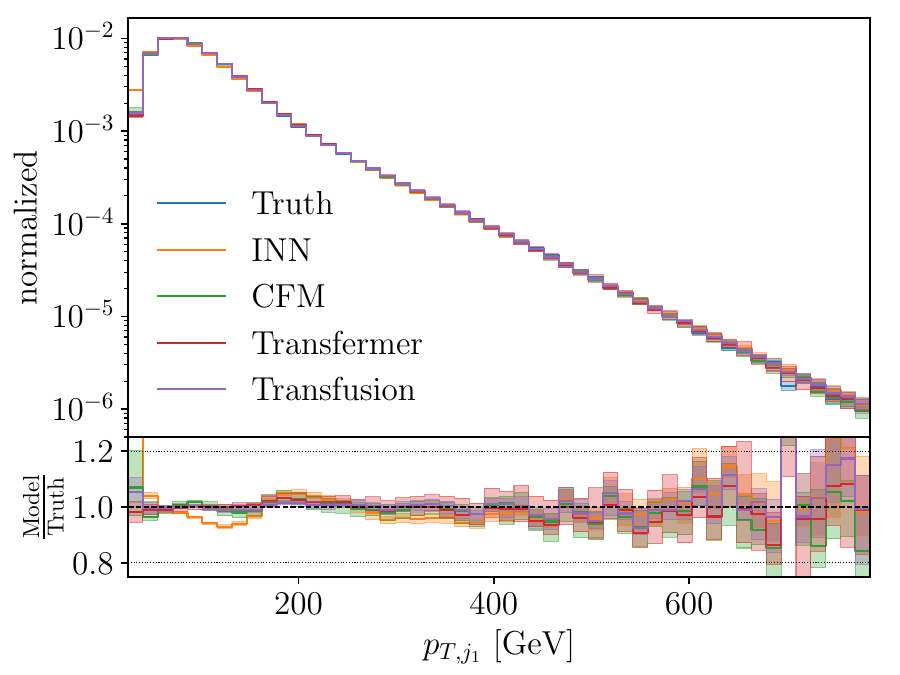} \\
    \includegraphics[width=0.49\textwidth,page=56]{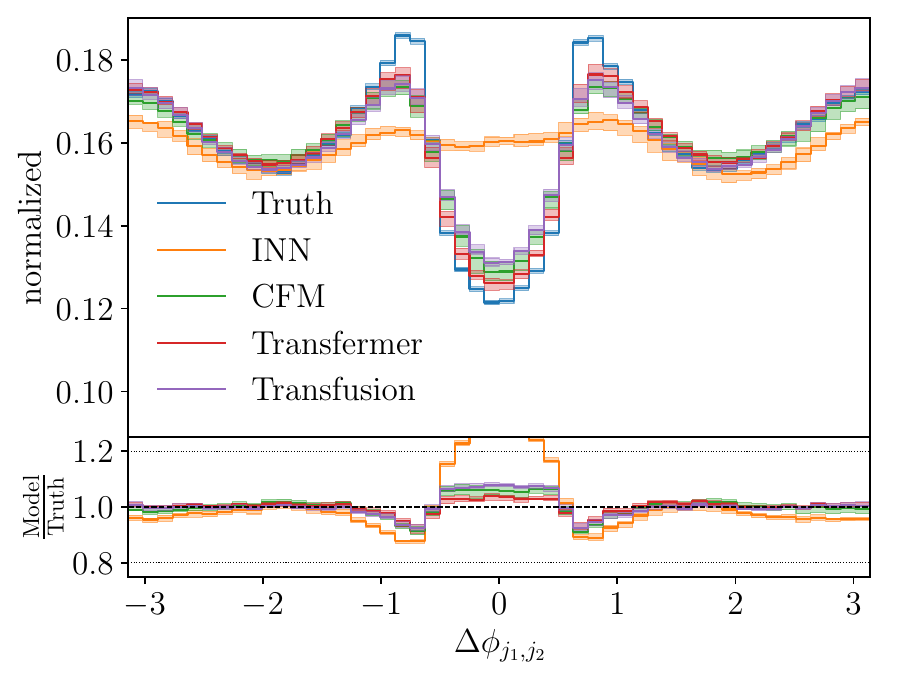}
    \includegraphics[width=0.49\textwidth,page=61]{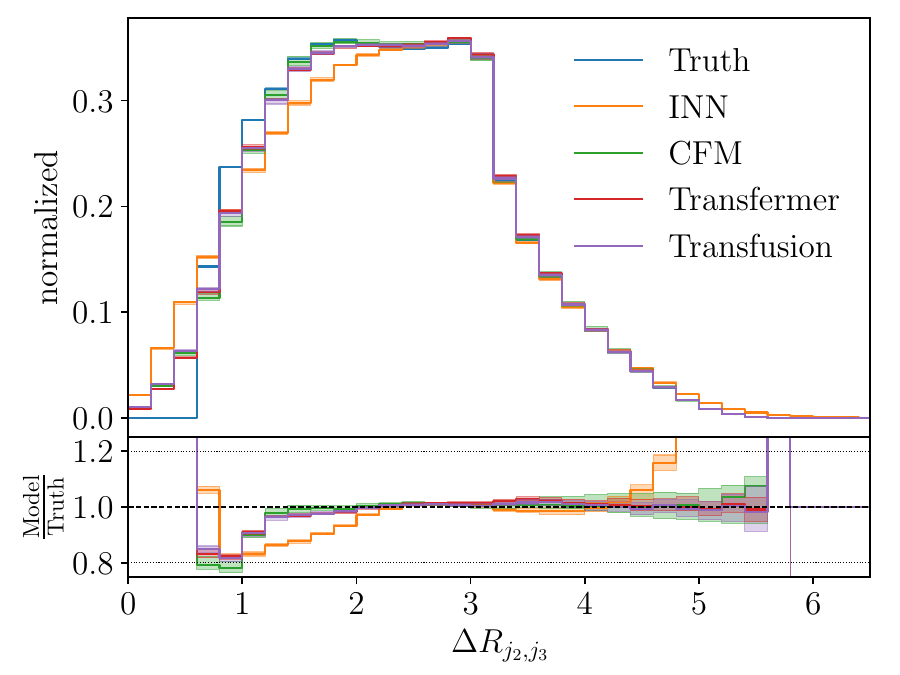}
    \caption{Reco-level distributions for different kinematic observables, obtained from the different generative transfer networks, conditioned on the hard-scattering momenta. Truth corresponds to the high-statistics training data.}
    \label{fig:observables}
\end{figure}

The matrix element method is an example of an LHC inference method, which is hugely attractive but only enabled by modern machine learning~\cite{Butter:2022vkj}. Specifically, it requires a fast and precise forward-transfer probability, an extremely efficient phase space mapping for the integration over the hard phase space, and a flexible encoding of the detector efficiency. We have shown, for a CP measurement in the associated production of a Higgs and a single top, how each of these tasks can be assigned to a neural network. This combination of three networks with modern architectures provides the required precision and speed.

To illustrate the performance of the different network architectures and MEM frameworks, we show a set of kinematic observables from the generative transfer networks in Fig.~\ref{fig:observables}.
Before integrating the likelihood, we can show the distributions at the reco-level and compare them to the truth, or training data.
We immediately see that the 
standard cINN is stable and extremely fast, but limited in its expressivity. The CFM diffusion network 
improves the performance significantly. The transformer architectures, \ie the cINN-driven Transfermer and the CFM-driven Transfusion 
deliver a precision at least on par with the CFM diffusion network.

For the likelihood, we have compared the extracted likelihoods from the different architectures with the hard-process target. As a benchmark, we first 
improved a range of numerical aspects of our concept paper~\cite{Butter:2022vkj}, with a focus on the integration with the Sampling-cINN. The improved 
precision of the integration raises two questions: a systematic bias in the minimum of the extracted 
likelihoods especially going from 400 to 10k events; and the optimality of the extracted
likelihoods seen in the widths in the CP-angle $\alpha$. These benchmark results are shown in Fig.~\ref{fig:inn_baseline}.

We then upgraded our two-network setup to a three-network setup,
with a learned acceptance as a function of phase space. In Fig.~\ref{fig:inn_baseline}, we saw that this removes the leading source of systematic bias, including the challenging SM-case $\alpha = 0^\circ$.

Next, we targeted the performance of the transfer network by replacing it with a more 
expressive, albeit slower CFM diffusion network. This did not improve the low-statistics results, but
for the high-statistics case of 10k events the Transfer-CFM showed a clear advantage over the cINN, as can be seen in Fig.~\ref{fig:cfm_acceptance}.

Finally, we solved the problem with the jet multiplicity of the cINN approach by applying a generative autoregressive Transfer-Transformer, \ie combining a transformer with a cINN network (Transfermer) and a CFM-model (Transfusion). In Fig.~\ref{fig:transfermer_fixed}, we saw that both transformer-based models outperformed the cINN, but showed similar performance as the Transfer-CFM. Notably, both transformer-based models can naturally be extended to describe a variable number of particles at both reco- and parton-level. This feature will eventually be needed for a proper description of the MEM at NLO.

In our LO example, all three models, CFM, Transfermer and Transfusion, parametrize the transfer probability flexibly and reliably. However, the Transfermer integration is approximately a factor 30 faster than the two diffusion-based models. This gap might eventually be closed using techniques like diffusion distillation~\cite{salimans2022progressive, song2023consistency, Buhmann:2023kdg}. Further improvements on the architecture, like the parallel Transfusion introduced in Appendix~\ref{sec:varjet_perminv}, might also improve the performance for more complex processes. Altogether, we conclude that a range of modern generative networks are available for the MEM, awaiting final judgment from an actual analysis. 

\section*{Acknowledgements}

We would like to thank the Baden-W\"urttemberg-Stiftung for funding through the program \textit{Internationale Spitzenforschung}, project
\textsl{Uncertainties --- Teaching AI its Limits}
(BWST\_IF2020-010). AB and TP are supported by the Deutsche
Forschungsgemeinschaft (DFG, German Research Foundation) under grant
396021762 -- TRR~257 \textsl{Particle Physics Phenomenology after the
Higgs Discovery}. 
AB, and NH are funded by the BMBF Junior Group Generative Precision Networks for Particle Physics (DLR 01IS22079).
TH is funded by the Carl-Zeiss-Stiftung through the
project \textsl{Model-Based AI: Physical Models and Deep Learning for
Imaging and Cancer Treatment}.
RW acknowledges support by FRS-FNRS (Belgian
National Scientific Research Fund) IISN projects 4.4503.16.
The authors acknowledge support by the state of Baden-Württemberg through bwHPC and the German Research Foundation (DFG) through grant no INST 39/963-1 FUGG (bwForCluster NEMO).
Computational resources have been provided by the
supercomputing facilities of the Université catholique de Louvain
(CISM/UCL) and the Consortium des Équipements de Calcul Intensif en
Fédération Wallonie Bruxelles (CÉCI) funded by the Fond de la
Recherche Scientifique de Belgique (F.R.S.-FNRS) under convention
2.5020.11 and by the Walloon Region. This work was supported by the 
DFG under Germany’s Excellence Strategy EXC 2181/1 - 390900948 
\textsl{The Heidelberg STRUCTURES Excellence Cluster}.

\clearpage
\appendix
\section{Network hyperparameters}

\begin{table}[ht!]
    \centering
    \begin{small} \begin{tabular}[t]{l|cc}
    \toprule
    Parameter & Acceptance & Multiplicities \\
    \midrule
    Optimizer & \multicolumn{2}{c}{Adam} \\
    Learning rate & \multicolumn{2}{c}{0.0001} \\
    LR schedule & \multicolumn{2}{c}{One-cycle} \\
    Maximum learning rate & \multicolumn{2}{c}{0.0003} \\
    Batch size & \multicolumn{2}{c}{1024} \\
    Epochs & \multicolumn{2}{c}{10} \\
    Number of layers & \multicolumn{2}{c}{6} \\
    Hidden nodes & \multicolumn{2}{c}{256} \\
    Activation function & \multicolumn{2}{c}{ReLU} \\
    Preprocessing & \multicolumn{2}{c}{$p_T, \eta, \phi, m$} \\
    Loss & Binary cross-entropy & Categorical cross-entropy \\
    Training samples & 5M & 3.4M \\
    Validation samples & 500k & 340k \\
    Testing samples & 4.5M & 3.1M \\
    Trainable parameters & 266k & 266k \\
    \bottomrule
    \end{tabular} \end{small}
    \caption{Hyperparameters of the classifiers learning the acceptance $\epsilon(x_\text{hard})$ (left) and the jet multiplicity used in Appendix~\ref{sec:varjet_perminv} (right).}
    \label{tab:classifier_hyperparams}
\end{table}

\begin{table}[ht!]
    \centering
    \begin{small} \begin{tabular}[t]{l|l}
    \toprule
    Parameter & Value \\
    \midrule
    Optimizer & Adam \\
    Learning rate & 0.001 \\
    LR schedule & Cosine-annealing \\
    Batch size & 16384 \\
    Epochs & 1000 \\
    Number of layers & 8 \\
    Feed-forward dimension & 512 \\
    Activation function & SiLU \\
    Training samples & 3.4M \\
    Validation samples & 340k \\
    Testing samples & 3.1M \\
    Trainable parameters & 3.2M \\
    ODE solver method& Runge-Kutta 4 \\
    Solver step-size & 0.05 \\
    \bottomrule
    \end{tabular}
    \hspace{0.5em}
    \begin{tabular}[t]{l|l}
    \toprule
    Parameter & Value \\
    \midrule
    Optimizer & RAdam \\
    Learning rate & 0.0001 \\
    LR schedule & One-cycle \\
    Maximum learning rate & 0.0003 \\
    Batch size & 1024 \\
    Epochs & 200 \\
    Number of heads & 8 \\
    Number of encoder layers & 6 \\
    Number of decoder layers & 8 \\
    Embedding dimension & 64 \\
    Transformer feed-forward dimension & 256 \\
    Number of subnet layers & 5 \\
    Subnet hidden nodes & 256 \\
    Subnet activation function & ReLU \\
    RQS spline bins & 16 \\
    Training samples & 3.4M \\
    Validation samples & 340k \\
    Testing samples & 3.1M \\
    Trainable parameters & 2.6M \\
    \bottomrule
    \end{tabular} \end{small}
    \caption{Hyperparameters of the CFM (left) and the Transfermer (right).}
    \label{tab:cfm_transfermer_hyperparams}
\end{table}

\begin{table}[ht!]
    \centering
    \begin{small} \begin{tabular}[t]{l|l}
    \toprule
    Parameter & Value \\
    \midrule
    Optimizer & Adam \\
    Learning rate & 0.001 \\
    LR schedule & Cosine-annealing \\
    Batch size & 8192 \\
    Epochs & 600 \\
    Number of heads & 8 \\
    Number of encoder layers & 6 \\
    Number of decoder layers & 8 \\
    Embedding dimension & 64 \\
    Transf. feed-forward dim & 256 \\
    Number of layers CFM & 6 \\
    Hidden nodes CFM & 400 \\
    Activation function CFM & ReLU \\
    Training samples & 3.4M \\
    Validation samples & 340k \\
    Testing samples & 3.1M \\
    Trainable parameters & 3.5M \\
    ODE solver method& Runge-Kutta, order 4 \\
    Solver step-size & 0.05 \\
    \bottomrule
    \end{tabular}
    \hspace{0.5em}
    \begin{tabular}[t]{l|l}
    \toprule
    Parameter & Value \\
    \midrule
    Optimizer & Adam \\
    Learning rate & 0.001 \\
    LR schedule & Cosine-annealing \\
    Batch size & 8192 \\
    Epochs & 600 \\
    Number of heads & 4 \\
    Number of encoder layers & 6 \\
    Number of decoder layers & 6 \\
    Embedding dimension & 128 \\
    Transf. feed-forward dim & 512 \\
    Training samples & 3.4M \\
    Validation samples & 340k \\
    Testing samples & 3.1M \\
    Trainable parameters & 2.7M \\
    ODE solver method & Runge-Kutta, order 4\\
    Solver step-size & 0.05 \\
    \bottomrule
    \end{tabular} \end{small}
    \caption{Hyperparameters of the autoregressive Transfusion (left) and the parallel Transfusion (right)}

    \label{tab:transfusion_hyperparams}
\end{table}

\section{Variable jet number and permutation invariance}
\label{sec:varjet_perminv}
\subsection*{Transfermer with variable jet number}

The Transfermer is easy to generalize to events with a variable number of jets at the reconstruction level. To this end, we split the inclusive transfer probability and evaluate it autoregressively,
\begin{align}
  p(x_\text{reco}, n|x_\text{hard})
  &= p(n|x_\text{hard}) \: p(x_\text{reco}|x_\text{hard}, n) \notag \\
  &= p(n|x_\text{hard}) \:
    p(x_\text{reco}^{(1:n_\text{min})}|x_\text{hard}, n)
    \prod_{i=n_\text{min}+1}^{n}
    p(x_\text{reco}^{(i)}|x_\text{reco}^{(1:i-1)},x_\text{hard}, n)
    \; ,
    \label{eq:varjet_transfer}
\end{align}
where $n$ is the number of final-state particles, $x_\text{reco}^{(1:k)}$ denotes the first $k$ reco-level
momenta, $x_\text{reco}^{(k)}$ denotes the $k$-th reco-level momentum
and $n_\text{min}$ is the minimal number of momenta for an accepted
event. The probability $p(n|x_\text{hard})$ can be extracted
using a simple classifier network with a categorical cross-entropy
loss and the number of additional jets as labels. The autoregressive factorization of $p(x_\text{reco}|x_\text{hard}, n)$ matches the way in which the Transfermer learns these probabilities. We pass the information about the number of additional jets to the Transfermer by appending it to the embedding of $x_\text{hard}$ in one-hot
encoded form. We can sample from the
transfer probability by first sampling the multiplicity
using the probabilities given by the classifier and then sampling the
momenta as described in Eq.\eqref{eq:transfermer_sampling}. Note that
it is even possible to generalize the Transfermer to a variable number
of hard-scattering momenta, because the transformer encoder accepts a variable number of inputs without any further changes to the architecture, making it a good candidate for a machine-learned MEM at NLO.

We train the jet multiplicity classifier with the hyperparameters given in Tab.~\ref{tab:classifier_hyperparams}. We observe that they are mostly flat for the top and Higgs, but there is a stronger variation as a function of the forward jet momentum, especially $\eta_j$. Like for the acceptance function, this is explained by ISR jets being tagged instead of the forward jet, leading to a lower probability of extra jets for $|\eta| > 2.4$.

\begin{figure}
    \includegraphics[width=0.33\textwidth,page=1]{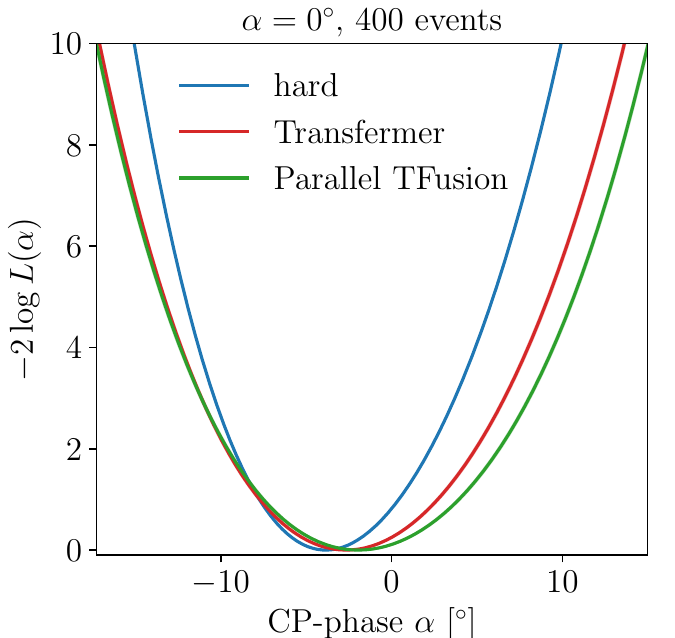}%
    \includegraphics[width=0.33\textwidth,page=2]{figs/likelihoods_appendix.pdf}%
    \includegraphics[width=0.33\textwidth,page=3]{figs/likelihoods_appendix.pdf}\\
    \includegraphics[width=0.33\textwidth,page=4]{figs/likelihoods_varjet.pdf}%
    \includegraphics[width=0.33\textwidth,page=5]{figs/likelihoods_varjet.pdf}%
    \includegraphics[width=0.33\textwidth,page=6]{figs/likelihoods_varjet.pdf}\\
    \includegraphics[width=0.33\textwidth,page=1]{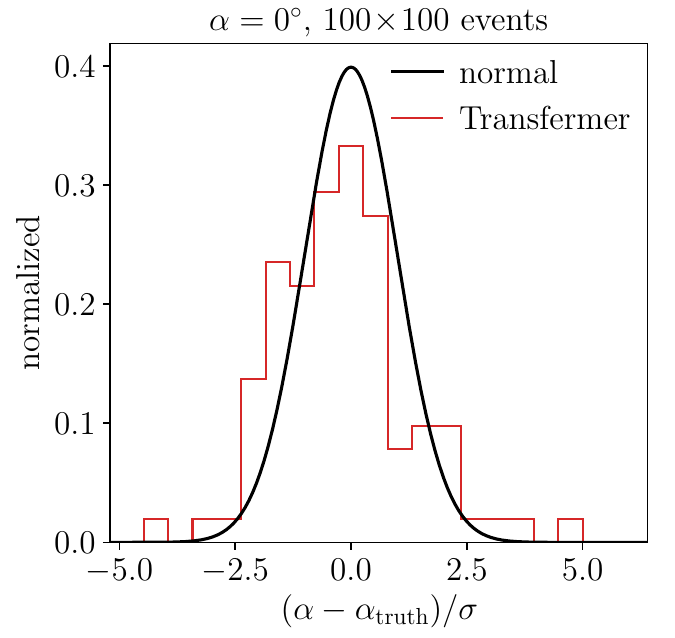}%
    \includegraphics[width=0.33\textwidth,page=2]{figs/pulls_varjet.pdf}%
    \includegraphics[width=0.33\textwidth,page=3]{figs/pulls_varjet.pdf}
    \caption{\textbf{Transfermer with variable jet numbers and parallel Transfusion:}
      likelihoods for different CP-angles using the Transfermer with variable jet multiplicity and the parallel Transfusion as the
      transfer probability. From top to bottom: likelihood for 400
      events, 10000 events, and pulls.}
    \label{fig:transfermer_varjet}
\end{figure}

We then run the MEM integration to obtain the results shown in
Fig.~\ref{fig:transfermer_varjet}. They are mostly similar to the
results with fixed multiplicity shown in
Fig.~\ref{fig:transfermer_fixed}. It shows that for our
specific process, we do not gain constraining power by including the
information from additional jets. However, that might be different for
other processes and especially at NLO. So the ability to deal with a
variable number of jets is still a valuable addition to our MEM toolbox.

\subsection*{Permutation-invariant Transfusion}

The Transfusion can be generalized to events with a variable jet number in complete analogy to the Transfermer. However, as diffusion models do not require invertibility, they allow for an additional approach in combining the transformer with the CFM network where we drop the autoregressive setup and instead generate all particle 4-momenta in parallel.\medskip

\begin{figure}[ht!]
    \centering
    \definecolor{Rcolor}{HTML}{E99595}
\definecolor{Gcolor}{HTML}{C5E0B4}
\definecolor{Gcolor_light}{HTML}{F1F8ED}
\definecolor{Bcolor}{HTML}{9DC3E6}
\definecolor{Ycolor}{HTML}{FFE699}
\definecolor{Ycolor_light}{HTML}{FFF7DE}

\tikzstyle{expr} = [rectangle, rounded corners=0.3ex, minimum width=1.5cm, minimum height=1cm, text centered, align=center, inner sep=0, fill=Ycolor, font=\large, draw]

\tikzstyle{small_cinn} = [double arrow, double arrow head extend=0cm, double arrow tip angle=130, inner sep=0, align=center, minimum width=1.1cm, minimum height=0.5cm, fill=Rcolor, draw]

\tikzstyle{small_cinn_black} = [small_cinn, minimum height=1.5cm, fill=black]

\tikzstyle{transformer} = [rectangle, rounded corners, minimum width=6cm, minimum height=2.4cm, font=\large, fill=Gcolor_light, draw]

\tikzstyle{attention} = [rectangle, rounded corners=0.3ex, minimum width=5.5cm, minimum height=1.2cm, align=center, fill=Gcolor, draw, font=\large]

\tikzstyle{txt_huge} = [align=center, font=\Huge, scale=2]
\tikzstyle{txt} = [align=center, font=\LARGE, minimum height=1cm]

\tikzstyle{arrow} = [thick,-{Latex[scale=1.0]}, line width=0.2mm, color=black]
\tikzstyle{line} = [thick, line width=0.2mm, color=black]

\begin{tikzpicture}[node distance=2cm, scale=0.65, every node/.style={transform shape}]

\node (hard1) [txt] {$x_{\text{hard}}^{(1)}$};
\node (hard2) [txt, right of=hard1, xshift=-0.5cm] {$...$};
\node (hard3) [txt, right of=hard2, xshift=-0.5cm] {$x_{\text{hard}}^{(3)}$};

\node (emb_hard1) [expr, below of=hard1, yshift=-0.5cm, rotate=90]{Emb};
\node (emb_hard2) [txt, below of=hard2, yshift=-0.5cm] {$...$};
\node (emb_hard3) [expr, below of=hard3,yshift=-0.5cm,  rotate=90]{Emb};

\node (TE) [transformer, below of=emb_hard2, yshift=-1.1cm, text width=4cm,
text depth=1.5cm, align=center] {Transformer-Encoder};
\node (TE_att) [attention, below of=TE, yshift=1.6cm] {Self-Attention: \\
Hard-level correlations};

\node (reco1) [txt, right of=hard3, xshift=1.75cm] {$x_{\text{reco}}^{(1)}(t|x_0^{(1)})$};
\node (reco3) [txt, right of=reco1] {$...$};
\node (reco6) [txt, right of=reco3] {$x_{\text{reco}}^{(n)}(t|x_0^{(n)})$};

\node (t) [txt, right of=reco6, xshift=0.5cm] {$t$};

\node (emb_reco1) [expr, below of=reco1, yshift=-0.5cm, rotate=90]{Emb};
\node (emb_reco3) [txt, below of=reco3, yshift=-0.5cm] {$...$};
\node (emb_reco6) [expr, below of=reco6, yshift=-0.5cm, rotate=90]{Emb};

\node (TD) [transformer, right of=TE, xshift=5.3cm, yshift=-0.8cm, text width=4cm,
text depth=3.1cm, align=center, minimum height=4cm] {Transformer-Decoder};
\node (TD_att) [attention, right of=TE_att, xshift=5.3cm] {Self-Attention: \\
Reco-level correlations};
\node (TD_crossatt) [attention, below of=TD_att, yshift=0.4cm] {Cross-Attention: \\
Combinatorics};

\node (inn1_b) [small_cinn_black, below of=reco1, yshift=-9.5cm, rotate=90]{cINN};
\node (inn6_b) [small_cinn_black, below of=reco6, yshift=-9.5cm, rotate=90]{cINN};
\node (inn1) [small_cinn, below of=reco1, yshift=-9.5cm, rotate=90]{CFM};
\node (inn3) [txt, below of=reco3, yshift=-9.5cm]{$...$};
\node (inn6) [small_cinn, below of=reco6, yshift=-9.5cm, rotate=90]{CFM};


\node (prob1) [txt, below of=inn1, yshift=-0.5cm]{$\Big( v^{(1)}(c^{(1)}, t ),$ };
\node (prob2) [txt, below of=inn3, yshift=-0.5cm]{$...$};
\node (prob6) [txt, below of=inn6, yshift=-0.5cm]{$, \; v^{(n)}(c^{(n)}, t ) \Big)$};
\node (prob) [txt, left of=prob1, xshift=-2.5 cm]{$v(x_\text{reco}(t), t | x_\text{hard}) = \;$};

\draw [arrow, color=black] (hard1.south) -- (emb_hard1.east);
\draw [arrow, color=black] (hard3.south) -- (emb_hard3.east);

\draw [arrow, color=black] (emb_hard1.west) -- (TE.north -| emb_hard1.west);
\draw [arrow, color=black] (emb_hard3.west) -- (TE.north -| emb_hard3.west);

\draw [arrow, color=black] (TE.south -| emb_hard1.west) --  ([yshift=-1cm]TE.south -| emb_hard1.west) -- ([yshift=-1cm]TE.south -| TD.west) ; 
\draw [arrow, color=black] (TE.south -| emb_hard3.west) --  ([yshift=-0.7cm]TE.south -| emb_hard3.west) -- ([yshift=-0.7cm]TE.south -| TD.west);

\draw [arrow, color=black] ([xshift=-0.2cm]reco1.south) -- ([xshift=-0.2cm]emb_reco1.east);
\draw [arrow, color=black] ([xshift=-0.2cm]reco6.south) -- ([xshift=-0.2cm]emb_reco6.east);

\draw [arrow, color=black] (emb_reco1.west) -- (TD.north -| emb_reco1.west);
\draw [arrow, color=black] (emb_reco6.west) -- (TD.north -| emb_reco6.west);

(A) (B);

\draw [arrow, color=black] ([xshift=-0.2cm]TD.south -| emb_reco1.west)  -- node [text width=1.5cm, pos=0.3, font=\LARGE, right] {$c^{(1)}$} ([xshift=-0.2cm]inn1.east -| emb_reco1.west);
\draw [arrow, color=black] ([xshift=-0.2cm]TD.south -| emb_reco6.west)  -- node [text width=1.5cm,pos=0.3, font=\LARGE, right] {$c^{(n)}$}  ([xshift=-0.2cm]inn6.east -| emb_reco6.west);

\draw [arrow, color=black] (inn1.west -| emb_reco1.west) -- (prob1.north -| emb_reco1.west);
\draw [arrow, color=black] (inn6.west -| emb_reco6.west) -- (prob6.north -| emb_reco6.west);

\draw [arrow, color=black] (t.south) --  ([yshift=-0.5cm]t.south) -- ([yshift=-0.5cm, xshift=0.2cm]t.south -| reco1.center) -- ([xshift=0.2cm]emb_reco1.east); 
\draw [arrow, color=black] ([yshift=-0.5cm, xshift=0.2cm]t.south -| reco6.center) -- ([xshift=0.2cm]emb_reco6.east); 
\draw [arrow, color=black] (t.south) --  ([yshift=-9.5cm]t.south) -- ([yshift=-9.5cm, xshift=0.2cm]t.south -| reco1.center) -- ([xshift=0.2cm]inn1.east); 
\draw [arrow, color=black] ([yshift=-9.5cm, xshift=0.2cm]t.south -| reco6.center) -- ([xshift=0.2cm]inn6.east); 

\end{tikzpicture}
    \caption{Parallel Transfusion architecture. Compared to the autoregressive setup we no longer use masked self-attention in the transformer decoder, but instead make it time-dependent.}
    \label{fig:transfusion_parallel_architecture}
\end{figure}
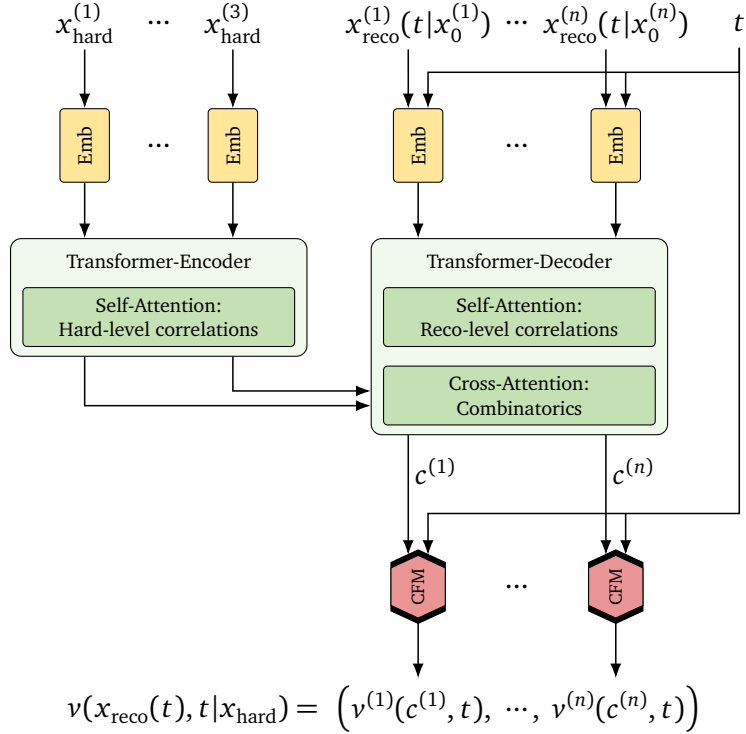

Before, in the autoregressive setup the transformer calculates a condition based on the hard-level momenta and the already generated reco-level momenta, which is then fed to the CFM that predicts the time-dependent velocity field. Crucially, the transformer itself has no time dependence. In the alternative parallel setup, the transformer decoder no longer sees the first $i-1$ reco-level particles $x_\text{reco}^{(1,\dots,i-1)}$ to describe $c^{(i)}$. Instead, its inputs are the conditional and time-dependent diffusion states $x_\text{reco}^{(1,\dots,n)}(t|x_0)$, as defined in Eq.\ref{eq:gaussian_probability_path_reparametrization}, of all $n$ reco-level particles, and the time t. The encoder, which acts only on the hard-level momente, is unchanged. Now, the transformer calculates time-dependent embeddings, one for each particle. These time-dependent embeddings are then again fed to a small CFM network predicting the velocity field. In this setup the velocity field of the $i^{\text{th}}$ particle is calculated as
\begin{align}
    v^{(i)}(c(e_\text{reco}(t), e_\text{hard}, t) , t) \; ,
    \label{eq:transfusionParallel_velocity}
\end{align}
where the transformer c is now a time-dependent function of the embeddings $e$ of all momenta. The overall setup is illustrated in Fig.~\ref{fig:transfusion_parallel_architecture}. In practice, a single linear layer is sufficient to map the transformer outputs to the velocity field components. Note, that during sampling the initial input to the transformer is the unconditional latent space vector $r$ which is then mapped onto $x_\text{reco}$ with the learned velocity field and the ODE solver. The parallel Transfusion setup naturally generalizes to varying particle multiplicities at both hard- and reco-level without requiring an arbitrary autoregressive order, as it is permutation-invariant at both levels.

\begin{figure}[t]
    \includegraphics[width=0.49\textwidth,page=23]{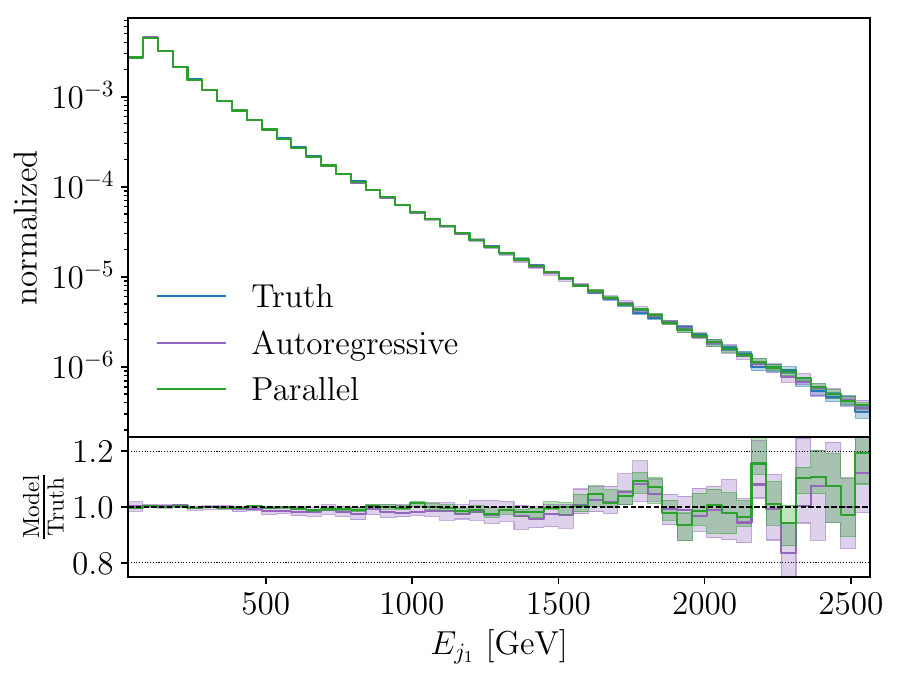}
    \includegraphics[width=0.49\textwidth,page=27]{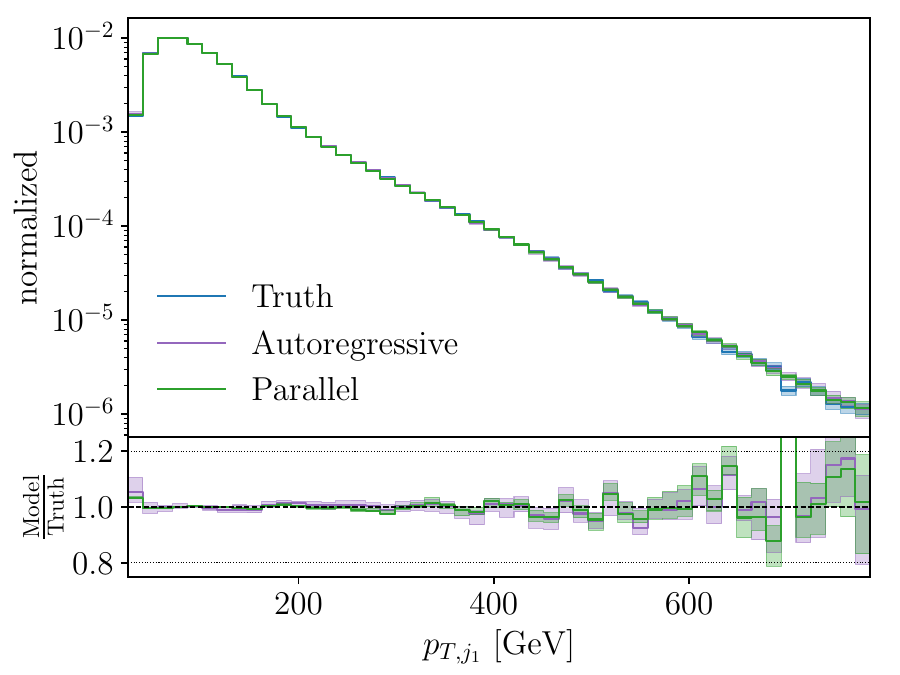} \\
    \includegraphics[width=0.49\textwidth,page=50]{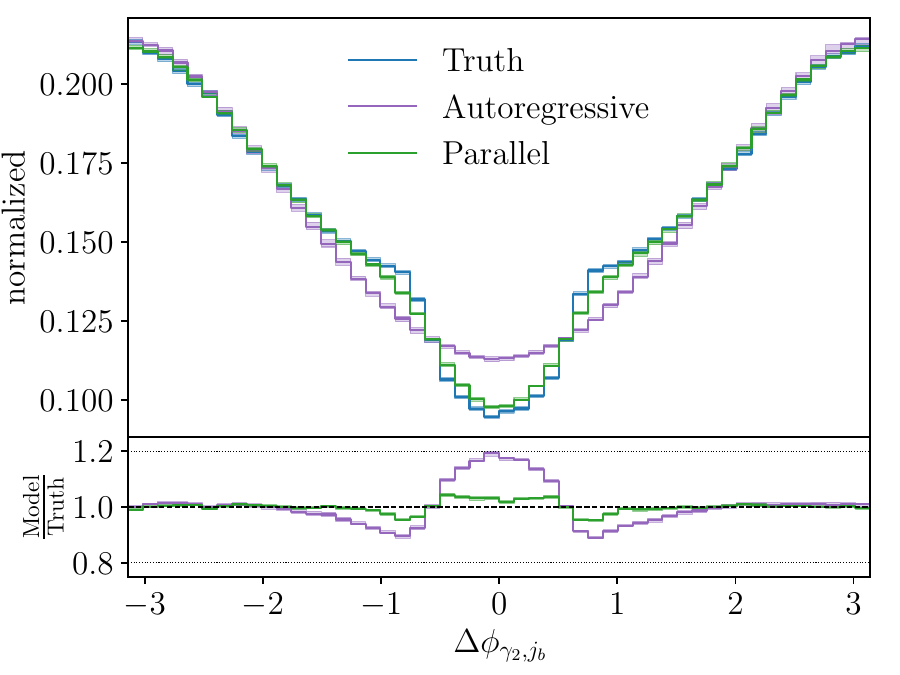}
    \includegraphics[width=0.49\textwidth,page=58]{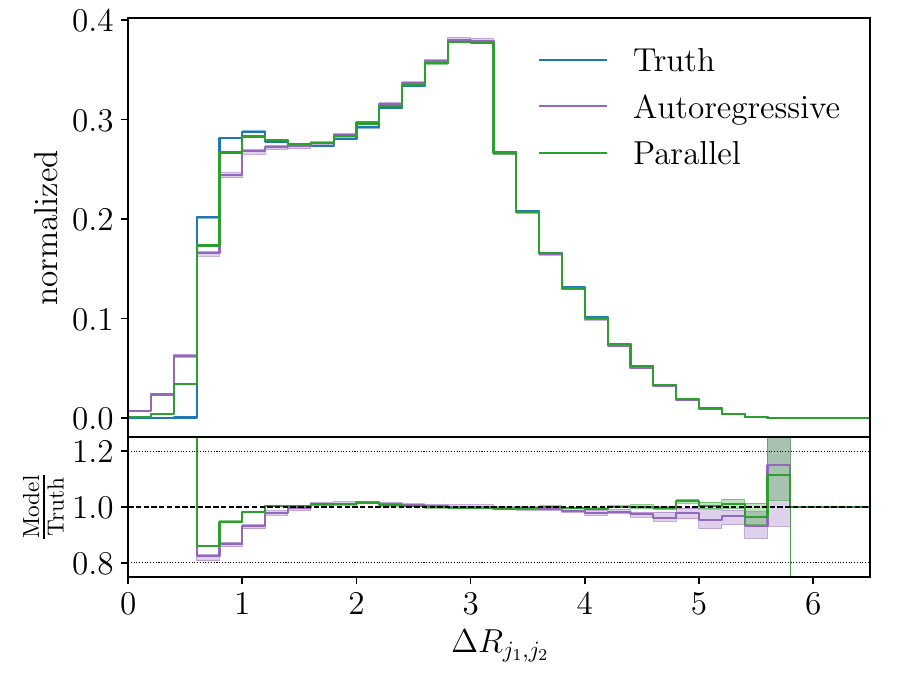}
    \caption{Reco-level distributions for different kinematic observables, obtained from the autoregressive and parallel Transfusion networks, conditioned on the hard-scattering momenta. Truth corresponds to the high-statistics training data.}
    \label{fig:observables_transfusion}
\end{figure}

Reco-level distributions for different kinematic observables are shown in Fig.~\ref{fig:observables_transfusion}. The marginal distributions show no difference between the parallel Transfusion and the other networks, but for the angular correlations we see the parallel Transfusion having a clear edge. Giving the transformer itself a time-dependence forces us to evaluate it repeatedly inside the ODE solver, making sampling and likelihood calculation in this setup even slower than for the pure CFM or the autoregressive Transfusion. We show integration results for 400 events using the parallel Transfusion in Fig.~\ref{fig:transfermer_varjet}, finding that they are comparable to the results from the autoregressive Transfusion. Due to the slow likelihood calculation this setup did not scale up to 10000 events.
The strong performance on the observable distribution level indicates that this architecture might proof useful in combination with speed-up techniques like diffusion distillation  or for applications that do not require likelihood calculation.

\section{Evaluating on Herwig}
\label{sec:herwig}
\begin{figure}
    \includegraphics[width=0.33\textwidth,page=1]{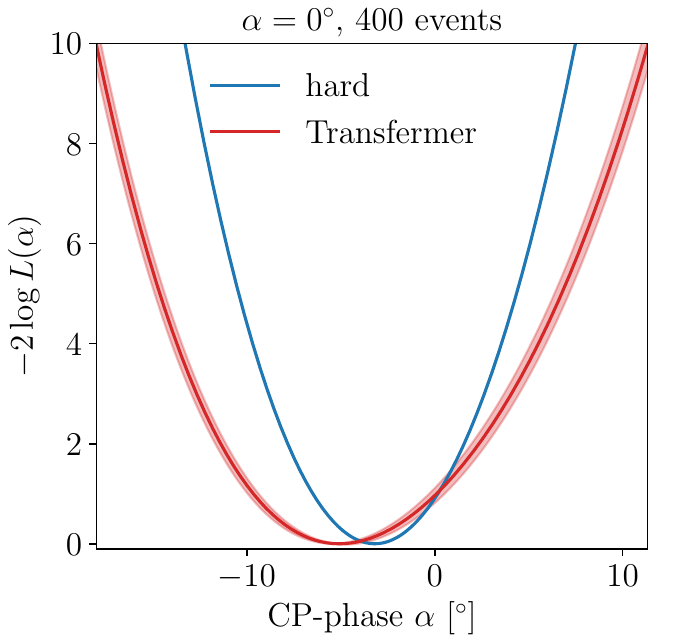}%
    \includegraphics[width=0.33\textwidth,page=2]{figs/likelihoods_herwig.pdf}%
    \includegraphics[width=0.33\textwidth,page=3]{figs/likelihoods_herwig.pdf}\\
    \includegraphics[width=0.33\textwidth,page=4]{figs/likelihoods_herwig.pdf}%
    \includegraphics[width=0.33\textwidth,page=5]{figs/likelihoods_herwig.pdf}%
    \includegraphics[width=0.33\textwidth,page=6]{figs/likelihoods_herwig.pdf}\\
    \includegraphics[width=0.33\textwidth,page=1]{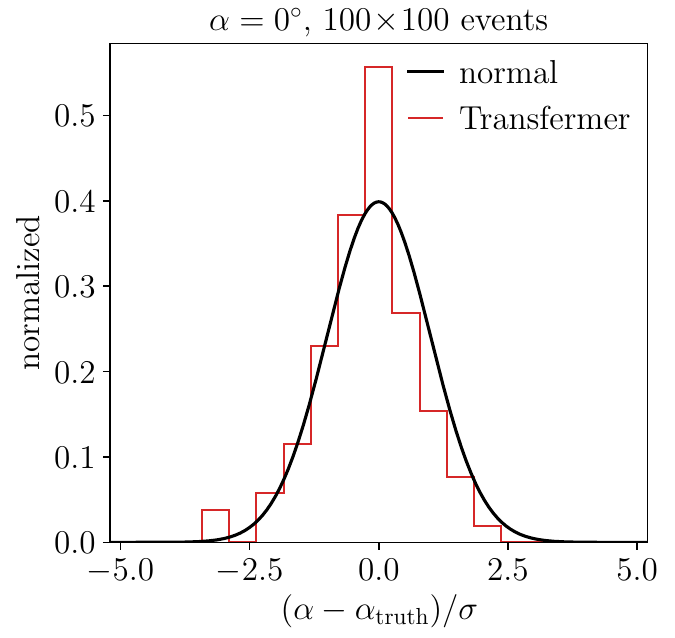}%
    \includegraphics[width=0.33\textwidth,page=2]{figs/pulls_herwig.pdf}%
    \includegraphics[width=0.33\textwidth,page=3]{figs/pulls_herwig.pdf}
    \caption{\textbf{Transfermer applied on \herwig simulation: }
      likelihoods for different CP-angles using the Transfermer trained on \pythia simulations but evaluated on \herwig simulations. From top to bottom: likelihood for 400
      events, 10000 events, and pulls.}
    \label{fig:transfermer_herwig}
\end{figure}

The critical backbone of our inference method is the learned transfer probability $\pmd (x_\text{reco}|x_\text{hard})$. We have demonstrated that generative networks can learn this conditional density from simulated data to very high precision. However, even a perfect network will only encode the forward transfer of the simulation, which is close but not necessarily identical to nature. In this section, we investigate how this impacts the results of our method by using different simulation setups: 
\begin{enumerate}
    \item a baseline simulation with \pythia for network training as described in Sec.~\ref{sec:intro};
    \item an alternative simulation based on \herwig~\cite{Bellm:2017bvx} for inference, emulating the truth reco-level data of the experiment. The detector effects are still modeled with \delphes. 
\end{enumerate} 
The results obtained with our method in this setup are shown in Fig.~\ref{fig:transfermer_herwig}. For 400 events we find that the extracted reco-level likelihoods mostly agree with the hard-level likelihoods. Note that the hard-level likelihoods are not fixed but also affected by the underlying simulation assumption, most visible for $\alpha=90^\circ$. This is because the fiducial hard-level likelihood is only defined on hard-level events $x_\text{hard}$ leading to accepted $x_\text{reco}$ events, which critically depends on the efficiency $\epsilon(x_\text{hard})$ of the underlying normalized transfer function $r$, as defined in Eqs.\eqref{eq:def_eff} and~\eqref{eq:p_fid_hard}. This effectively encodes a dependence on the assumed forward simulation
\begin{align}
    p_\text{fid}(x_\text{hard}|\alpha)\equiv p_{\pythia}(x_\text{hard}|\alpha)\;.
\end{align}
Hence, evaluating the fiducial hard-level likelihoods on the \herwig simulation can generally lead to a bias in the likelihood distribution. In the high-statistics scenario with 10k events, we observe good agreement for $\alpha=0, \; 45^\circ$, comparable to the results when evaluating on \pythia, which means we can assume 
\begin{align}
p_{\pythia}(x_\text{hard}|\alpha)\approx p_{\herwig}(x_\text{hard}|\alpha)\;.
\label{eq:pythi_herwig_sim}
\end{align}
In these cases, we find that the reco-level likelihood obtained using our method still agrees well with the hard-scattering likelihood, and the results are still well-calibrated. However, for $\alpha=90^\circ$ the hard-level likelihoods are significantly off from the true value, indicating that Eq.\eqref{eq:pythi_herwig_sim} is no longer valid. Further, training the transfer function on events that do not follow the true distribution of the measured data may introduce $\alpha$-dependent effects. Consequently, we also find a large deviation between the hard- and reco-level likelihoods.

\clearpage
\bibliography{tilman,generative,literature} 
\end{document}